\documentclass[reprint,showpacs,preprintnumbers,pre,superscriptaddress]{revtex4-1}
\usepackage[T1]{fontenc}
\usepackage[latin9]{inputenc}
\usepackage{mathrsfs}
\usepackage{amsmath}
\usepackage{amssymb}
\usepackage{graphicx}
\usepackage{wasysym}

\makeatletter
\usepackage{bm}\usepackage{amsfonts}
\usepackage{color}

\makeatother

\begin{document}
\title{Geodesic path for the minimal energy cost in shortcuts to isothermality}
\author{Geng Li}
\affiliation{Graduate School of China Academy of Engineering Physics, Beijing 100193,
China}
\author{Jin-Fu Chen}
\affiliation{Beijing Computational Science Research Center, Beijing 100193, China}
\affiliation{Graduate School of China Academy of Engineering Physics, Beijing 100193,
China}
\author{C. P. Sun}
\affiliation{Graduate School of China Academy of Engineering Physics, Beijing 100193,
China}
\affiliation{Beijing Computational Science Research Center, Beijing 100193, China}
\author{Hui Dong}
\email{hdong@gscaep.ac.cn}

\affiliation{Graduate School of China Academy of Engineering Physics, Beijing 100193,
China}
\begin{abstract}
Shortcut to isothermality is a driving strategy to steer the system
to its equilibrium states within finite time, and enables evaluating
the impact of a control promptly. Finding optimal scheme to minimize
the energy cost is of critical importance in applications of this
strategy in pharmaceutical drug test, biological selection, and quantum
computation. We prove the equivalence between designing the optimal
scheme and finding the geodesic path in the space of control parameters.
Such equivalence allows a systematic and universal approach to find
the optimal control to reduce the energy cost. We demonstrate the
current method with examples of a Brownian particle trapped in controllable
harmonic potentials.
\end{abstract}
\maketitle
\emph{Introduction.}-- Boosting system to its steady state is critical
to promptly evaluate the impact of a control \cite{Nichol2015,Ahmad2017,Ogbunugafor2016,Iram2020,Ilker2021,Albash2018,Takahashi2019,GueryOdelin2019}.
In biological systems, the quest to timely evaluate the impact of
therapy or genotypes posts a requirement to steer the system to reach
its steady state with a considerable tunable rate \cite{Nichol2015,Ahmad2017,Ogbunugafor2016,Iram2020,Ilker2021}.
In adiabatic quantum computation, the task of solving the optimization
problem is converted to the problem of driving systems from a trivial
ground state to another nontrivial ground state. The speedup of the
computational process needs to steer the system to the target ground
state in finite time \cite{Albash2018,Takahashi2019,GueryOdelin2019}.
These quests to tune the system within finite time while keep it in
equilibrium is eagerly needed.

Shortcut to isothermality was proposed as a finite-time driving strategy
to steer the system evolving along the path of instantaneous equilibrium
states \cite{Li2017}. The strategy has been applied in reducing transition
time between equilibrium states \cite{Albay2019,Albay2020,Albay2020a},
improving the efficiency of free-energy estimation \cite{Li2021},
constructing finite-time heat engines \cite{Pancotti2020,Nakamura2020,Plata2020},
and controlling biological evolutions \cite{Iram2020,Ilker2021}.
The cost of the finite-time operation is the additional energy cost
due to irreversibility posted by the fundamental thermodynamic law.
Minimizing such cost is in turn relevant to optimize the heat engine
\cite{Schmiedl2007a,Schmiedl2007,Tu2014} and reconstruct the energy
landscape of biological macromolecules \cite{Atilgan2004,Ytreberg2006,Vaikuntanathan2008}.
A question arises naturally, how to find the optimal control protocol
to minimize the irreversible energy cost in shortcuts to isothermality.

\begin{figure}[!htp]
\includegraphics{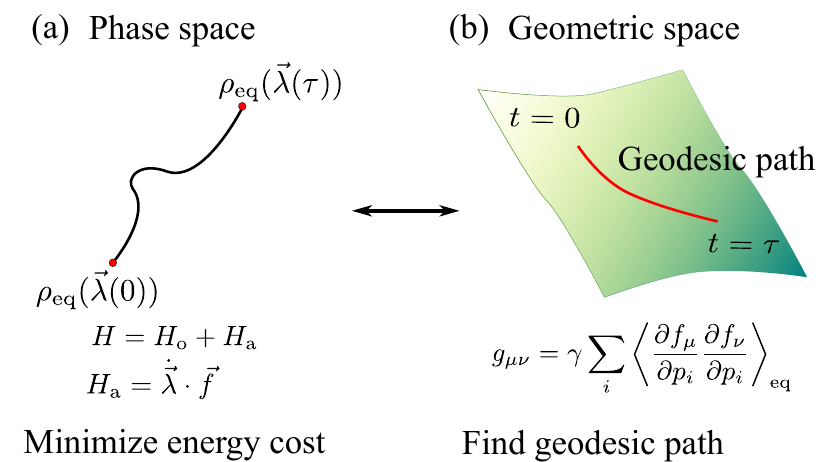} \caption{(Color online) The equivalence between designing the optimal control
protocol and finding the geodesic path in the parametric space. (a)
The evolution of the system controlled by the shortcut scheme. An
auxiliary Hamiltonian $H_{\mathrm{a}}=\dot{\vec{\lambda}}\cdot\vec{f}$
is added to steer the evolution along the instantaneous equilibrium
state $\rho_{\mathrm{eq}}(\vec{\lambda}(t))$ of the original Hamiltonian
$H_{\mathrm{o}}$. Designing the optimal control protocol normally
requires to minimize the energy cost in the shortcut scheme. (b) The
geodesic path in the equivalent geometric space. We can convert the
designing task into finding the geodesic path in the geometric space
with the metric $g_{\mu\nu}=\gamma\sum_{i}\langle\partial_{p_{i}}f_{\mu}\partial_{p_{i}}f_{\nu}\rangle_{\mathrm{eq}}$.}
\label{fig1}
\end{figure}

In this Letter, we present a systematic approach for finding the optimal
protocol to minimize the energy cost. In Fig. \ref{fig1}, we show
the equivalence of designing the optimal control to finding the geodesic
path on a Riemannian manifold, spanned by the control parameters \cite{Salamon1983,Nielsen2006,Crooks2007,Sivak2012,Chen2021}.
In turn, the powerful tools developed in geometry are adapted for
solving the optimal control protocol. Our scheme is exemplified with
a single Brownian particle in the harmonic potential with controllable
stiffness and central position.

\emph{Geometric approach} -- The system is described by the Hamiltonian
$H_{\mathrm{o}}(\vec{x},\vec{p},\vec{\lambda})=\sum_{i}p_{i}^{2}/2+U_{\mathrm{o}}(\vec{x},\vec{p},\vec{\lambda})$
with the coordinate $\vec{x}\equiv(x_{1},x_{2},\cdots,x_{N})$ and
the momentum $\vec{p}\equiv(p_{1},p_{2},\cdots,p_{N})$. It is immersed
in a thermal reservoir with a constant temperature $T$. $\vec{\lambda}(t)\equiv(\lambda_{1},\lambda_{2},\cdots,\lambda_{M})$
are time-dependent control parameters. For simplicity, we have set
the mass of the system as a unit. In the shortcut scheme, an auxiliary
Hamiltonian $H_{\mathrm{a}}(\vec{x},\vec{p},t)$ is added to steer
the system evolving along the instantaneous equilibrium states of
the original Hamiltonian $H_{\mathrm{o}}$ in the finite-time interval
$\tau$ with boundary conditions $H_{\mathrm{a}}(0)=H_{\mathrm{a}}(\tau)=0$.
The dynamical evolution under the total Hamiltonian $H=H_{\mathrm{o}}+H_{\mathrm{a}}$
is described by the Langevin equation as

\begin{align}
\dot{x}_{i} & =\frac{\partial H}{\partial p_{i}},\nonumber \\
\dot{p}_{i} & =-\frac{\partial H}{\partial x_{i}}-\gamma\dot{x}_{i}+\xi_{i}(t),\label{eq:ulaneq}
\end{align}
where $\gamma$ is the dissipation rate and $\vec{\xi}\equiv(\xi_{1},\xi_{2},\cdots,\xi_{N})$
are random variables of the Gaussian white noise. The evolution equation
of the system distribution $\rho(\vec{x},\vec{p},t)=\delta(\vec{x}-\vec{x}(t))\delta(\vec{p}-\vec{p}(t))$
for a trajectory $\text{[}\vec{x}(t),\vec{p}(t)]$ is described by
the Liouville equation as $\text{\ensuremath{\partial_{t}\rho}}=-\sum_{i}[\partial_{x_{i}}(\dot{x}_{i}\rho)+\partial_{p_{i}}(\dot{p_{i}}\rho)].$
By averaging over different noise realizations $[\vec{\xi}(t)]$,
we obtain the evolution of the observable probability distribution
$P(\vec{x},\vec{p},t)\equiv\langle\rho(\vec{x},\vec{p},t)\rangle_{\vec{\xi}}=\iint D[\vec{x}(t)]D[\vec{p}(t)]\mathscr{T}[\vec{x}(t),\vec{p}(t)]\delta(\vec{x}-\vec{x}(t))\delta(\vec{p}-\vec{p}(t))$
as \cite{Supple}
\begin{equation}
\frac{\partial P}{\partial t}=\sum_{i}[-\frac{\partial}{\partial x_{i}}(\frac{\partial H}{\partial p_{i}}P)+\frac{\partial}{\partial p_{i}}(\frac{\partial H}{\partial x_{i}}P+\gamma\frac{\partial H}{\partial p_{i}}P)+\frac{\gamma}{\beta}\frac{\partial^{2}P}{\partial p_{i}^{2}}],\label{eq:kramerseq}
\end{equation}
where $\beta\equiv1/(k_{\mathrm{B}}T)$ is the inverse temperature
with the Boltzmann constant $k_{\mathrm{B}}$. Here $\mathscr{T}[\vec{x}(t),\vec{p}(t)]$
is the probability of the trajectory $[\vec{x}(t),\vec{p}(t)]$ associated
with a noise realization $[\vec{\xi}(t)]$ \cite{Reichl1998}. To
ensure the instantaneous equilibrium distribution 
\begin{equation}
P(\vec{x},\vec{p},t)=P_{\mathrm{eq}}(\vec{x},\vec{p},\vec{\lambda})=\mathrm{e}^{\beta[F(\vec{\lambda})-H_{\mathrm{o}}(\vec{x},\vec{p},\vec{\lambda})]},\label{eq:insequdis}
\end{equation}
the auxiliary Hamiltonian is proved \cite{Li2017} to have the form
$H_{\mathrm{a}}(\vec{x},\vec{p},t)=\dot{\vec{\lambda}}\cdot\vec{f}(\vec{x},\vec{p},\vec{\lambda})$
with $\vec{f}(\vec{x},\vec{p},\vec{\lambda})$ satisfying 
\begin{equation}
\sum_{i}[\frac{\gamma}{\beta}\frac{\partial^{2}f_{\mu}}{\partial p_{i}^{2}}-\gamma p_{i}\frac{\partial f_{\mu}}{\partial p_{i}}+\frac{\partial f_{\mu}}{\partial p_{i}}\frac{\partial U_{\mathrm{o}}}{\partial x_{i}}-p_{i}\frac{\partial f_{\mu}}{\partial x_{i}}]=\frac{dF}{d\lambda_{\mu}}-\frac{\partial U_{\mathrm{o}}}{\partial\lambda_{\mu}},\label{eq:additionalHamiltonianeq}
\end{equation}
where $F\equiv-\beta^{-1}\ln[\iint d\vec{x}d\vec{p}\exp(-\beta H_{\mathrm{o}})]$
is the free energy. The boundary conditions are presented explicitly
as $\dot{\vec{\lambda}}(0)=\dot{\vec{\lambda}}(\tau)=0$.

The cost of the energy in the shortcut scheme is evaluated by the
average work $W\equiv\left\langle \int_{0}^{\tau}dt\partial_{t}H\right\rangle _{\vec{\xi}}$
\cite{Jarzynski1997,Sekimoto2010,Seifert2012,Li2019}, explicitly
as 
\begin{equation}
W=\Delta F+\gamma\sum_{i}\varint_{0}^{\tau}dt\iint d\vec{x}d\vec{p}\left(\frac{\partial H_{\mathrm{a}}}{\partial p_{i}}\right)^{2}P_{\mathrm{eq}},\label{eq:meanwork}
\end{equation}
where $\Delta F=F(\vec{\lambda}(\tau))-F(\vec{\lambda}(0))$ is the
free energy difference. Detailed derivation of Eq. (\ref{eq:meanwork})
is presented in the supplementary materials \cite{Supple}. To consider
the finite-time effect, we define the irreversible work $W_{\mathrm{irr}}\equiv W-\Delta F,$
which follows 
\begin{equation}
W_{\mathrm{irr}}=\gamma\sum_{\mu\nu i}\varint_{0}^{\tau}dt\dot{\lambda}_{\mu}\dot{\lambda}_{\nu}\left\langle \frac{\partial f_{\mu}}{\partial p_{i}}\frac{\partial f_{\nu}}{\partial p_{i}}\right\rangle _{\mathrm{eq}},\label{eq:girrevcost}
\end{equation}
with $\left\langle \cdot\right\rangle _{\mathrm{eq}}=\iint d\vec{x}d\vec{p\;}[\cdot]P_{\mathrm{eq}}$.
It follows from Eq. (\ref{eq:girrevcost}) that the integrand scales
as $\tau^{-2}$ through reducing the time $s\equiv t/\tau$, which
results in the $1/\tau$ scaling \cite{Salamon1983} of the irreversible
work, i.e, $W_{\mathrm{irr}}\propto1/\tau$. Such a $1/\tau$ scaling,
predicted in various finite-time studies \cite{Curzon1975,Broeck2005,Schmiedl2007,Esposito2010,Tu2012,Wang2012,Tomas2012,Ryabov2016,Cavina2017,Ma2018a,Ma2018},
was recently verified for the ideal gas system \cite{Ma2020} at the
long-time limit. It is worth noting that in the shortcut scheme the
current scaling is valid for any duration time $\tau$ with no requirement
of the long-time limit \cite{Salamon1983,Cavina2017,Scandi2019,Crooks2007,Chen2021a}.

In the space of the control parameters $\vec{\lambda}$, we define
a positive semi-definite metric 
\begin{equation}
g_{\mu\nu}=\gamma\text{\ensuremath{\sum_{i}}}\left\langle \frac{\partial f_{\mu}}{\partial p_{i}}\frac{\partial f_{\nu}}{\partial p_{i}}\right\rangle _{\mathrm{eq}},\label{eq:gmetric}
\end{equation}
whose positive semi-definiteness is proved in the supplementary materials
\cite{Supple}. With this metric, the length of a curve in the current
geometric space is characterized via the thermodynamic length \cite{Salamon1983,Crooks2007,Sivak2012,Scandi2019,Chen2021}
as $\mathcal{L}=\int_{0}^{\tau}dt\sum_{\mu\nu}\sqrt{\dot{\lambda}_{\mu}\dot{\lambda}_{\nu}g_{\mu\nu}}$,
which provides a lower bound of the irreversible work $W_{\mathrm{irr}}$
as 
\begin{equation}
W_{\mathrm{irr}}\geq\frac{\mathcal{L}^{2}}{\tau}.\label{eq:cauchyschwarz}
\end{equation}
The lower bound is reached with the optimal control scheme $\vec{\lambda}(t)$
$\text{(}0<t<\tau)$, determined by the geodesic equation 
\begin{equation}
\ddot{\lambda}_{\mu}+\sum_{\nu\kappa}\Gamma_{\nu\kappa}^{\mu}\dot{\lambda}_{\nu}\dot{\lambda}_{\kappa}=0,\label{eq:geodesiceq}
\end{equation}
with the given boundary conditions $\vec{\lambda}(0)$ and $\vec{\lambda}(\tau)$.
Here the Christoffel symbol is defined as $\Gamma_{\nu\kappa}^{\mu}\equiv\frac{1}{2}\sum_{\iota}(g^{-1})_{\iota\mu}(\partial_{\lambda_{\kappa}}g_{\iota\nu}+\partial_{\lambda_{\nu}}g_{\iota\kappa}-\partial_{\lambda_{\iota}}g_{\nu\kappa})$.
For the case with the single control parameter $\lambda(t),$ the
analytical solution \cite{Sivak2012} for Eq. (\ref{eq:geodesiceq})
is obtained as $\dot{\lambda}(t)=(\lambda(\tau)-\lambda(0))g(\lambda(t))^{-1}/\int_{0}^{\tau}dt'g(\lambda(t'))^{-1}$,
with $g=\gamma\langle(\partial_{p}f)^{2}\rangle$. For the case with
multiple parameters, the shooting method is an available option which
treats the two-point boundary-value problem as an initial-value problem
\cite{Berger2007}. See the supplementary materials for details about
the shooting method to our problems \cite{Supple}.

The strategy of current formalism is shown in Fig. \ref{fig1}. Firstly,
we obtain the control operators $\vec{f}(\vec{x},\vec{p},\vec{\lambda})$
in Fig. \ref{fig1}(a) by solving Eq. (\ref{eq:additionalHamiltonianeq}).
Secondly, the metric $g_{\mu\nu}$ in Fig. \ref{fig1}(b) for the
parametric space is calculated via Eq. (\ref{eq:gmetric}). Finally,
the optimal control is obtained by solving the geodesic equation in
Eq. (\ref{eq:geodesiceq}). The current strategy provides an effective
approach to find the optimal control to minimize the energy cost,
i.e., the total work done during the shortcut-to-isothermal process.
The strategy is illustrated through two examples with one or two control
parameters as follows.

\textit{Brownian motion in the harmonic potential}-- The Brownian
particle is trapped by the one-dimensional breathing harmonic potential
with tunable stiffness $\lambda(t)$ under the Hamiltonian $H_{\mathrm{o}}(x,p,\lambda)=p^{2}/2+\lambda(t)x^{2}/2$.
Its auxiliary Hamiltonian was derived in Ref. \cite{Li2017} as $H_{\mathrm{a}}(x,p,t)=\dot{\lambda}f(x,p,\lambda)$
with $f=1/(4\gamma\lambda)[(p-\gamma x)^{2}+\lambda x^{2}]$. The
metric in Eq. (\ref{eq:gmetric}) in this case reduces to \cite{Supple}
\begin{equation}
g=\frac{\lambda+\gamma^{2}}{4\gamma\beta\lambda^{3}}.\label{eq:oharsing}
\end{equation}
And the lower bound of the irreversible work is reached by the protocol
satisfying the geodesic equation $\ddot{\lambda}+\dot{\lambda}^{2}\partial_{\lambda}g/(2g)=0$.
The solution 
\begin{equation}
\lambda_{\mathrm{gp}}(t)=\frac{\sqrt{1+2\gamma^{2}(m_{\mathrm{s}}-n_{\mathrm{s}}t/\tau)}+1}{2(m_{\mathrm{s}}-n_{\mathrm{s}}t/\tau)},\label{eq:singleopt}
\end{equation}
offers an optimal protocol to minimize the energy cost in the shortcut
scheme. Here $m_{\mathrm{s}}=1/\lambda(0)+\gamma^{2}/(2\lambda(0))$
and $n_{\mathrm{s}}=1/\lambda(0)+\gamma^{2}/(2\lambda(0))-1/\lambda(\tau)-\gamma^{2}/(2\lambda(\tau))$
are constants for single control-parameter case. And the irreversible
work of the geodesic protocol reaches its minimum $W_{\mathrm{irr}}^{\mathrm{min}}=\int_{0}^{\tau}\dot{\lambda}^{2}gdt=n_{\mathrm{s}}^{2}/\tau$,
which is consistent with the lower bound given by the thermodynamic
length $\mathcal{L}=\int_{0}^{\tau}\sqrt{\dot{\lambda}^{2}g}dt=n_{\mathrm{s}}$
through the relation $W_{\mathrm{irr}}^{\mathrm{min}}=\mathcal{L}^{2}/\tau$.

\textit{Underdamped Brownian motion with two control parameters} --We
consider a Brownian particle moving in the one-dimensional harmonic
potential with Hamiltonian $H_{\mathrm{o}}(x,p,\lambda)=p^{2}/2+\lambda_{1}x^{2}/2-\lambda_{2}x$.
The auxiliary Hamiltonian for the shortcut scheme takes the form \cite{Supple}
$H_{\mathrm{a}}(x,p,t)=\sum_{\mu=1}^{2}\dot{\lambda}_{\mu}f_{\mu}(x,p,\lambda_{1},\lambda_{2})$
with 
\begin{align}
f_{1} & =\frac{\left(p-\gamma x\right)^{2}+\lambda_{1}x^{2}}{4\gamma\lambda_{1}}-\frac{\lambda_{2}p}{2\lambda_{1}^{2}}+(\frac{\gamma\lambda_{2}}{2\lambda_{1}^{2}}-\frac{\lambda_{2}}{2\gamma\lambda_{1}})x,\nonumber \\
f_{2} & =\frac{p}{\lambda_{1}}-\frac{\gamma x}{\lambda_{1}}.\label{eq:f1aux}
\end{align}
The metric in Eq. (\ref{eq:gmetric}) for the control parameters $\vec{\lambda}$
is obtained as
\begin{equation}
g=\left(\begin{array}{cc}
\frac{1}{4\beta\gamma\lambda_{1}^{2}}+\frac{\gamma}{4\beta\lambda_{1}^{3}}+\frac{\gamma\lambda_{2}^{2}}{\lambda_{1}^{4}} & -\frac{\gamma\lambda_{2}}{\lambda_{1}^{3}}\\
-\frac{\gamma\lambda_{2}}{\lambda_{1}^{3}} & \frac{\gamma}{\lambda_{1}^{2}}
\end{array}\right).
\end{equation}
The geodesic equation follows
\begin{eqnarray}
 &  & \ddot{\lambda}_{1}-\frac{\dot{\lambda}_{1}^{2}(3\gamma^{2}+2\lambda_{1})}{2\lambda_{1}(\gamma^{2}+\lambda_{1})}=0,\nonumber \\
 &  & \ddot{\lambda}_{2}-\frac{2\dot{\lambda}_{1}\dot{\lambda}_{2}}{\lambda_{1}}+\frac{\dot{\lambda}_{1}^{2}\lambda_{2}(\gamma^{2}+2\lambda_{1})}{2\lambda_{1}^{2}(\gamma^{2}+\lambda_{1})}=0,\label{eq:twohargeo}
\end{eqnarray}
with the boundary conditions $\vec{\lambda}(0)$ and $\vec{\lambda}(\tau)$
.

\begin{figure}[!htp]
\includegraphics[width=8.5cm]{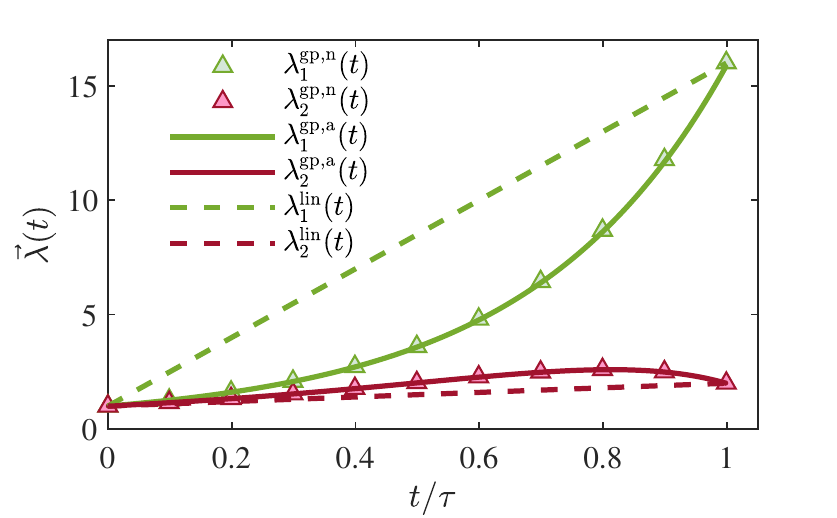} \caption{(Color online) Geodesic protocols for the control with two parameters.
In the simulation, we have set the temperature and the dissipation
rate as $k_{\mathrm{B}}T=1$ and $\gamma=1$. The parameters change
from the initial point $\vec{\lambda}(0)$=($1,1$) to the final point
$\vec{\lambda}(\tau)$=($16,2$). The triangles represent the numerical
geodesic protocol $\vec{\lambda}^{\mathrm{gp,n}}(t)$ while the solid
lines represent the analytical geodesic protocol $\vec{\lambda}^{\mathrm{gp,a}}(t)$.
The dash lines represent the linear protocol $\vec{\lambda}^{\mathrm{lin}}(t)$.
The numerical geodesic protocol (triangles) coincides well with the
analytical geodesic protocol (solid lines).}
\label{fig2}
\end{figure}

The optimal scheme can be obtained by solving equations above using
a general numerical method, i.e., the shooting method \cite{Berger2007}.
Here we firstly solve these equations numerically to provide a general
perspective on our scheme. With the initial point $\vec{\lambda}(0)$,
we choose an initial rate $\dot{\vec{\lambda}}(0+)$ and solve the
geodesic equation with the Eular algorithm to obtain a trial solution
$\vec{\lambda}^{\mathrm{tri}}(\tau)$. Newton's method is utilized
for updating the rate $\dot{\vec{\lambda}}(0+)$ to reduce the distance
between the trial solution $\vec{\lambda}^{\mathrm{tri}}(\tau)$ and
the target point $\vec{\lambda}(\tau)$. In the simulation, we have
chosen the parameters $\vec{\lambda}(0)=(1,1)$, $\vec{\lambda}(\tau)=(16,2)$,
$k_{\mathrm{B}}T=1$, and $\gamma=1$. The geodesic path for the optimal
control is illustrated as $\vec{\lambda}^{\mathrm{gp,n}}(t)$ (triangles)
in Fig. \ref{fig2}.

Fortunately, the analytical geodesic protocol for Eq.~(\ref{eq:twohargeo})
can be obtained as
\begin{align}
\dot{\lambda}_{1} & =\frac{w_{\mathrm{b}}}{\tau}\sqrt{\frac{\lambda_{1}^{3}}{\lambda_{1}+\gamma^{2}}},\nonumber \\
\frac{\lambda_{2}}{\lambda_{1}} & =m_{\mathrm{b}}t/\tau+n_{\mathrm{b}},\label{eq:analyticalg}
\end{align}
where $w_{\mathrm{b}}=-[2\sqrt{1+\gamma^{2}/\lambda_{1}}+\ln(\sqrt{1+\gamma^{2}/\lambda_{1}}-1)-\ln(\sqrt{1+\gamma^{2}/\lambda_{1}}+1)]|_{\lambda_{1}(0)}^{\lambda_{1}(\tau)}$,
$m_{\mathrm{b}}=(\lambda_{2}(\tau)\lambda_{1}(0)-\lambda_{2}(0)\lambda_{1}(\tau))/(\lambda_{1}(\tau)\lambda_{1}(0))$,
and $n_{\mathrm{b}}=\lambda_{2}(0)/\lambda_{1}(0)$ are constants.
In Fig. \ref{fig2}, we show the match between the optimal control
obtained from the numerical calculation $\vec{\lambda}^{\mathrm{gp,n}}(t)$
(triangles) and the analytical solution $\vec{\lambda}^{\mathrm{gp,a}}(t)$
(solid lines). For the comparison, we also show the protocol of the
simple linear control $\vec{\lambda}^{\mathrm{lin}}(t)=(\vec{\lambda}(\tau)-\vec{\lambda}(0))t/\tau+\vec{\lambda}(0)$.

\begin{figure}[!htp]
\includegraphics[width=8.5cm]{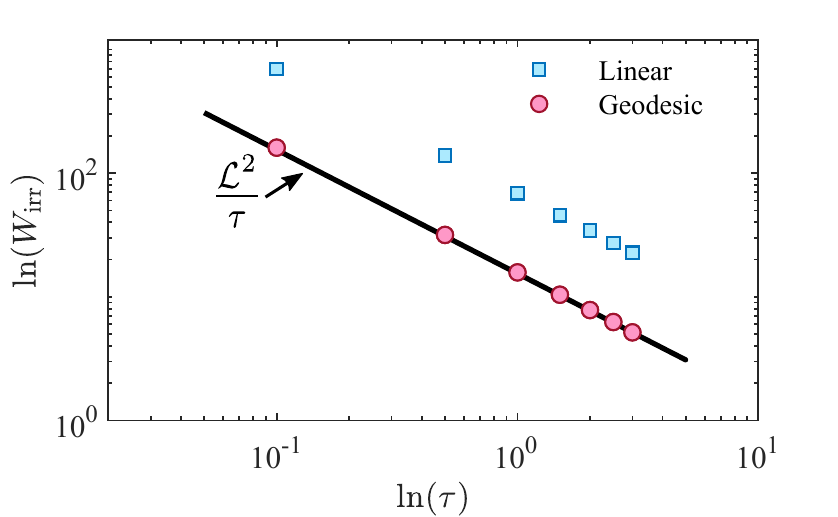} \caption{(Color online) The irreversible work of the geodesic protocol (red
circles) and the linear protocol (blue squares). The black line represents
the theoretical lower bound given by the thermodynamic length, i.e.,
Eq. (\ref{eq:cauchyschwarz}). We perform the simulation for different
control duration $\tau\in\{0.1,0.5,1.0,1.5,2.0,2.5,3.0\}$. The irreversible
cost given by the geodesic protocol is lower than that from the linear
protocol. And the lower bound given by the geodesic protocol matches
the one given by the thermodynamic length.}
\label{fig3}
\end{figure}

To validate our results of optimization, we calculate the irreversible
work for the single Brownian particle in the controllable harmonic
potential with two control parameters by solving the Langevin equation
(\ref{eq:ulaneq}) through the Euler algorithm \cite{Supple,Frenkel2001}.
The average work is calculated by the ensemble average of the stochastic
work over $10^{5}$ stochastic trajectories. Details of the simulation
are presented in the supplementary material \cite{Supple}. In Fig.
\ref{fig3}, we plot the irreversible work $W_{\mathrm{irr}}$ as
a function of duration $\tau\in\{0.1,0.5,1.0,1.5,2.0,2.5,3.0\}$ for
both the geodesic path (red circles) and the simple linear control
(blue squares). The geodesic protocol results in a lower irreversible
work than that from the linear protocol. The black line shows the
analytical results $W_{\mathrm{irr}}^{\mathrm{min}}=\mathcal{L}^{2}/\tau$,
where the thermodynamic length $\mathcal{L}$ is calculated as $\mathcal{L}=\int_{0}^{\tau}dt\sum_{\mu\nu}\sqrt{\dot{\lambda}_{\mu}\dot{\lambda}_{\nu}g_{\mu\nu}}=\sqrt{w_{\mathrm{b}}^{2}/(4\beta\gamma)+\gamma m_{\mathrm{b}}^{2}}$.
The simulation results match the lower bound presented by the thermodynamic
length, illustrated by the coincide of the simulated results (red
circles) with the theoretical line (black line). Figure \ref{fig3}
shows that the geodesic protocol can largely reduce the irreversible
work, which therefore proves our findings about the geometric property
of the control-parameter space in the shortcut scheme. Our findings
simplify the procedure of finding the optimal control protocol in
the shortcut scheme by applying the tools of Riemannian geometry.

\emph{Conclusions.}-- In summary, we have provided a geometric approach
to find the optimal control scheme to steer the evolution of the system
along the path of instantaneous equilibrium states to reduce the energy
cost. The proven equivalence between designing the optimal control
and finding the geodesic path in the parametric space allows the application
of the methods developed in Riemannian geometry to solve the optimization
problem in thermodynamics. We have applied our approach into the Brownian
particle system tuned by both one and two control parameters to find
the optimal control for reducing energy cost. Analytical and numerical
results have verified that the geodesic protocol can largely reduce
the irreversible work in the shortcut scheme. Our strategy shall provide
an effective tool to design the optimal finite-time control with the
lowest energy cost.

Our results demonstrate that the optimal control with the minimal
energy cost to transfer the system between equilibrium states is to
steer the system evolving along the geodesic path. Once the initial
and final equilibrium states are given, the geodesic path is determined
by the geodesic equation (\ref{eq:geodesiceq}) for the given system.
The dynamics of the system is covered by the metric $g_{\mu\nu}$
in Eq. (\ref{eq:gmetric}) without the need to treat the system on
a case-by-case basis. An intuitive determination of the performance
of the controls is allowed with the proportional relation in Eq. (\ref{eq:cauchyschwarz})
between the minimal energy cost and the square of the length of the
geodesic path.

\emph{Acknowledgement.}--This work is supported by the National Natural
Science Foundation of China (NSFC) (Grants No. 12088101, No. 11534002,
No. 11875049, No. U1930402, No. U1930403 and No. 12047549) and the
National Basic Research Program of China (Grant No. 2016YFA0301201).

\bibliographystyle{apsrev4-1}
\bibliography{ref}

\end{document}


\title{Supplementary materials: Geodesic path for the minimal energy cost
in shortcuts to isothermality}
\author{Geng Li}
\affiliation{Graduate School of China Academy of Engineering Physics, Beijing 100193,
China}
\author{Jin-Fu Chen}
\affiliation{Beijing Computational Science Research Center, Beijing 100193, China}
\affiliation{Graduate School of China Academy of Engineering Physics, Beijing 100193,
China}
\author{C. P. Sun}
\affiliation{Graduate School of China Academy of Engineering Physics, Beijing 100193,
China}
\affiliation{Beijing Computational Science Research Center, Beijing 100193, China}
\author{Hui Dong}
\email{hdong@gscaep.ac.cn}

\affiliation{Graduate School of China Academy of Engineering Physics, Beijing 100193,
China}

\maketitle
The supplementary materials are devoted to provide detailed derivations
in the main context.

\tableofcontents{}

\section{The modified Kramers equation\label{SSec-one}}

The Kramers equation \citep{Reichl1998} was developed for describing
systems with the form of Hamiltonian $H_{\mathrm{o}}=\vec{p}^{2}/2+U_{\mathrm{o}}(\vec{x},\vec{\lambda})$.
In the main text, we consider systems controlled by shortcuts to isothermality
with the form of Hamiltonian $H=H_{\mathrm{o}}+H_{\mathrm{a}}$, with
boundary conditions $H_{\mathrm{a}}(0)=H_{\mathrm{a}}(\tau)=0$ at
the initial time $t=0$ and the final time $t=\tau$. In this section,
we derive a modified Kramers equation \citep{Li2017} for the total
Hamiltonian $H$.

The evolution equation of the system probability distribution $\text{\ensuremath{\rho(\vec{x},\vec{p},t)=\delta(\vec{x}-\vec{x}(t))\delta(\vec{p}-\vec{p}(t))}}$
for a trajectory $\text{[}\vec{x}(t),\vec{p}(t)]$ is governed by
the Liouville equation as 
\begin{equation}
\frac{\partial\rho}{\partial t}=-\sum_{i}[\frac{\partial}{\partial x_{i}}(\dot{x}_{i}\rho)+\frac{\partial}{\partial p_{i}}(\dot{p_{i}}\rho)],\label{seq:oevoequ}
\end{equation}
where the evolution of variables is described by the Langevin equation
as 
\begin{align}
\dot{x}_{i} & =\frac{\partial H}{\partial p_{i}},\nonumber \\
\dot{p}_{i} & =-\frac{\partial H}{\partial x_{i}}-\gamma\dot{x}_{i}+\xi_{i}(t).\label{seq:ulaneq}
\end{align}
With Eqs.~(\ref{seq:oevoequ}) and~(\ref{seq:ulaneq}), we obtain
\begin{equation}
\frac{\partial\rho}{\partial t}=\sum_{i}[-\frac{\partial}{\partial x_{i}}(\frac{\partial H}{\partial p_{i}}\rho)+\frac{\partial}{\partial p_{i}}(\frac{\partial H}{\partial x_{i}}\rho+\gamma\frac{\partial H}{\partial p_{i}}\rho)-\frac{\partial}{\partial p_{i}}(\xi_{i}\rho)].\label{seq:ranevo}
\end{equation}
An observable probability to describe the average effect of the probability
distribution over different realizations of $[\vec{\xi}(t)]$ can
be defined as 
\begin{eqnarray}
P(\vec{x},\vec{p},t) & \equiv & \langle\rho(\vec{x},\vec{p},t)\rangle_{\vec{\xi}}\nonumber \\
 & = & \text{\ensuremath{\iint}}D[\vec{x}(t)]D[\vec{p}(t)]\text{\ensuremath{\mathscr{T}}}[\vec{x}(t),\vec{p}(t)]\rho(\vec{x},\vec{p},t)\nonumber \\
 & = & \iint D[\vec{x}(t)]D[\vec{p}(t)]\mathscr{T}[\vec{x}(t),\vec{p}(t)]\delta(\vec{x}-\vec{x}(t))\delta(\vec{p}-\vec{p}(t)),
\end{eqnarray}
where $\mathscr{T}[\vec{x}(t),\vec{p}(t)]$ is the probability of
the trajectory $[\vec{x}(t),\vec{p}(t)]$ associated with a noise
realization $[\vec{\xi}(t)]$ \citep{Reichl1998}. In the following,
we derive the evolution equation for the observable probability $P(\vec{x},\vec{p},t)$.

Equation~(\ref{seq:ranevo}) can be rewritten as 
\begin{equation}
\frac{\partial\rho}{\partial t}=-\hat{L}_{\mathrm{d}}\rho-\hat{L}_{\mathrm{s}}\rho,\label{seq:reoranevo}
\end{equation}
with the deterministic operator 
\[
\hat{L}_{\mathrm{d}}(t)\equiv\sum_{i}[\frac{\partial}{\partial x_{i}}(\frac{\partial H}{\partial p_{i}})-\frac{\partial}{\partial p_{i}}(\frac{\partial H}{\partial x_{i}}+\gamma\frac{\partial H}{\partial p_{i}})],
\]
and the stochastic operator 
\[
\hat{L}_{\mathrm{s}}(t)\equiv\sum_{i}\text{\ensuremath{\xi_{i}}}\frac{\partial}{\partial p_{i}}.
\]
Introduce a new probability $\phi(\vec{x},\vec{p},t)$ that satisfies
\begin{equation}
\rho(\vec{x},\vec{p},t)=\text{\ensuremath{\mathcal{T}}}e^{-\int_{0}^{t}\hat{L}_{\mathrm{d}}(t')dt'}\phi(\vec{x},\vec{p},t)=\hat{R}(t)\phi(\vec{x},\vec{p},t),\label{seq:newpro}
\end{equation}
where $\hat{R}(t)\equiv\text{\ensuremath{\mathcal{T}}}e^{-\int_{0}^{t}\hat{L}_{\mathrm{d}}(t')dt'}$
with $\mathcal{\mathcal{T}}$ representing the time-order operator.
Substituting Eq. (\ref{seq:newpro}) into Eq.~(\ref{seq:reoranevo}),
we obtain 
\begin{equation}
\frac{\partial\phi}{\partial t}=-\hat{O}\phi,\label{seq:newevo}
\end{equation}
where $\hat{O}(t)\equiv\hat{R}^{-1}(t)\hat{L}_{\mathrm{s}}(t)\hat{R}(t)$
with $\hat{R}^{-1}(t)\equiv(\text{\ensuremath{\mathcal{T}}}e^{-\int_{0}^{t}\hat{L}_{\mathrm{d}}(t')dt'})^{-1}$.
Equation~(\ref{seq:newevo}) has a formal solution as 
\begin{eqnarray}
\phi(\vec{x},\vec{p},t) & = & \mathcal{\mathcal{T}}e^{-\int_{0}^{t}\hat{O}(t')dt'}\phi(\vec{x},\vec{p},0)\nonumber \\
 & = & [\sum_{n}(-1)^{n}\int_{0}^{t}dt_{n}\int_{0}^{t_{n}}dt_{n-1}\cdots\int_{0}^{t_{2}}dt_{1}\hat{O}(t_{n})\hat{O}(t_{n-1})\cdots\hat{O}(t_{1})]\phi(\vec{x},\vec{p},0).\label{seq:newevofs}
\end{eqnarray}
Averaging it over different realizations of $[\vec{\xi}]$, we derive
that 
\begin{eqnarray}
\langle\phi(\vec{x},\vec{p},t)\rangle_{\vec{\xi}} & = & [\sum_{n}(-1)^{n}\int_{0}^{t}dt_{n}\int_{0}^{t_{n}}dt_{n-1}\cdots\int_{0}^{t_{2}}dt_{1}\langle\hat{O}(t_{n})\hat{O}(t_{n-1})\cdots\hat{O}(t_{1})\rangle_{\vec{\xi}}]\phi(\vec{x},\vec{p},0)\nonumber \\
 & = & [\sum_{n}(-1)^{2n}\int_{0}^{t}dt_{2n}\int_{0}^{t_{2n}}dt_{2n-1}\cdots\int_{0}^{t_{2}}dt_{1}\langle\hat{O}(t_{2n})\hat{O}(t_{2n-1})\cdots\hat{O}(t_{1})\rangle_{\vec{\xi}}]\phi(\vec{x},\vec{p},0).\label{seq:averagephi}
\end{eqnarray}
In the second step, we have considered the fact that the noise $\vec{\xi}$
is Gaussian satisfying $\langle\xi_{i}(t)\rangle_{\vec{\xi}}=0$ and
$\langle\xi_{i}(t)\xi_{j}(t')\rangle_{\vec{\xi}}=2\gamma k_{\mathrm{B}}T\delta_{ij}\delta(t-t')$.
The higher-order moments that contain odd number of $\xi_{i}$ are
zero. The remaining terms containing even number of $\xi_{i}$ can
be decomposed into a sum of products of the second order moment $\langle\xi_{i}(t)\xi_{j}(t')\rangle_{\vec{\xi}}$.
For example, the fourth order moment follows 
\begin{eqnarray}
\langle\xi_{i}(t_{1})\xi_{j}(t_{2})\xi_{k}(t_{3})\xi_{l}(t_{4})\rangle_{\vec{\xi}} & = & \langle\xi_{i}(t_{1})\xi_{j}(t_{2})\rangle_{\vec{\xi}}\langle\xi_{k}(t_{3})\xi_{l}(t_{4})\rangle_{\vec{\xi}}+\langle\xi_{i}(t_{1})\xi_{k}(t_{3})\rangle_{\vec{\xi}}\langle\xi_{j}(t_{2})\xi_{l}(t_{4})\rangle_{\vec{\xi}}\nonumber \\
 &  & +\langle\xi_{i}(t_{1})\xi_{l}(t_{4})\rangle_{\vec{\xi}}\langle\xi_{j}(t_{2})\xi_{k}(t_{3})\rangle_{\vec{\xi}}.\label{seq:fourthmoment}
\end{eqnarray}
In the right hand side of Eq. (\ref{seq:fourthmoment}), the second
and third terms vanish because of the time order $t\geq t_{2n}\cdots\geq t_{1}$.
Therefore, Eq.~(\ref{seq:averagephi}) reduces to the form

\begin{eqnarray}
\langle\phi(\vec{x},\vec{p},t)\rangle_{\vec{\xi}} & = & [\sum_{n}(\int_{0}^{t}dt_{2n}\int_{0}^{t_{2n}}dt_{2n-1}\langle\hat{O}(t_{2n})\hat{O}(t_{2n-1})\rangle_{\vec{\xi}})(\int_{0}^{t_{2n-1}}dt_{2n-2}\int_{0}^{t_{2n-2}}dt_{2n-3}\langle\hat{O}(t_{2n-2})\hat{O}(t_{2n-3})\rangle_{\vec{\xi}})\nonumber \\
 &  & \times\cdots(\int_{0}^{t_{3}}dt_{2}\int_{0}^{t_{2}}dt_{1}\langle\hat{O}(t_{2})\hat{O}(t_{1})\rangle)]\phi(\vec{x},\vec{p},0).\label{seq:avenewpro}
\end{eqnarray}
One of the integral in Eq.~(\ref{seq:avenewpro}) is calculated as
\begin{align}
 & \int_{0}^{t_{3}}dt_{2}\int_{0}^{t_{2}}dt_{1}\langle\hat{O}(t_{2})\hat{O}(t_{1})\rangle\nonumber \\
 & =\sum_{ij}\int_{0}^{t_{3}}dt_{2}\int_{0}^{t_{2}}dt_{1}\langle\xi_{i}(t_{2})\xi_{j}(t_{1})\rangle_{\vec{\xi}}\hat{R}^{-1}(t_{2})\frac{\partial}{\partial p_{i}}\hat{R}(t_{2})\hat{R}^{-1}(t_{1})\frac{\partial}{\partial p_{j}}\hat{R}(t_{1})\nonumber \\
 & =2\gamma k_{\mathrm{B}}T\sum_{ij}\int_{0}^{t_{3}}dt_{2}\int_{0}^{t_{2}}dt_{1}\delta_{ij}\delta(t_{2}-t_{1})\hat{R}^{-1}(t_{2})\frac{\partial}{\partial p_{i}}\hat{R}(t_{2})\hat{R}^{-1}(t_{1})\frac{\partial}{\partial p_{j}}\hat{R}(t_{1})\nonumber \\
 & =\gamma k_{\mathrm{B}}T\sum_{i}\int_{0}^{t_{3}}dt_{2}\hat{R}^{-1}(t_{2})\frac{\partial^{2}}{\partial p_{i}^{2}}\hat{R}(t_{2}).\label{seq:exponphi}
\end{align}
Then Eq.~(\ref{seq:avenewpro}) proceeds as 
\begin{eqnarray}
\langle\phi(\vec{x},\vec{p},t)\rangle_{\vec{\xi}} & = & [\sum_{n}(\gamma k_{\mathrm{B}}T)^{n}(\sum_{i}\int_{0}^{t}dt_{2n}\hat{R}^{-1}(t_{2n})\frac{\partial^{2}}{\partial p_{i}^{2}}\hat{R}(t_{2n}))(\sum_{i}\int_{0}^{t_{2n}}dt_{2n-2}\hat{R}^{-1}(t_{2n-2})\frac{\partial^{2}}{\partial p_{i}^{2}}\hat{R}(t_{2n-2}))\nonumber \\
 &  & \times\cdots(\sum_{i}\int_{0}^{t_{4}}dt_{2}\hat{R}^{-1}(t_{2})\frac{\partial^{2}}{\partial p_{i}^{2}}\hat{R}(t_{2}))]\phi(\vec{x},\vec{p},0)\nonumber \\
 & = & [\sum_{n}(\gamma k_{\mathrm{B}}T)^{n}(\sum_{i}\int_{0}^{t}dt_{n}\hat{R}^{-1}(t_{n})\frac{\partial^{2}}{\partial p_{i}^{2}}\hat{R}(t_{n}))(\sum_{i}\int_{0}^{t_{n}}dt_{n-1}\hat{R}^{-1}(t_{n-1})\frac{\partial^{2}}{\partial p_{i}^{2}}\hat{R}(t_{n-1}))\nonumber \\
 &  & \times\cdots(\sum_{i}\int_{0}^{t_{2}}dt_{1}\hat{R}^{-1}(t_{1})\frac{\partial^{2}}{\partial p_{i}^{2}}\hat{R}(t_{1}))]\phi(\vec{x},\vec{p},0).\label{seq:finalavep}
\end{eqnarray}
Taking derivative of Eq. (\ref{seq:finalavep}) over time $t$, we
obtain 
\begin{eqnarray}
\frac{\partial}{\partial t}\langle\phi(\vec{x},\vec{p},t)\rangle_{\vec{\xi}} & = & [\sum_{n=0}^{\infty}(\gamma k_{\mathrm{B}}T)^{n}(\sum_{i}\hat{R}^{-1}(t)\frac{\partial^{2}}{\partial p_{i}^{2}}\hat{R}(t))(\sum_{i}\int_{0}^{t}dt_{n-1}\hat{R}^{-1}(t_{n-1})\frac{\partial^{2}}{\partial p_{i}^{2}}\hat{R}(t_{n-1}))\nonumber \\
 &  & \times\cdots(\sum_{i}\int_{0}^{t_{2}}dt_{1}\hat{R}^{-1}(t_{1})\frac{\partial^{2}}{\partial p_{i}^{2}}\hat{R}(t_{1}))]\phi(\vec{x},\vec{p},0)\nonumber \\
 & = & \gamma k_{\mathrm{B}}T(\sum_{i}\hat{R}^{-1}(t)\frac{\partial^{2}}{\partial p_{i}^{2}}\hat{R}(t))[\sum_{n=1}^{\infty}(\gamma k_{\mathrm{B}}T)^{n-1}(\sum_{i}\int_{0}^{t}dt_{n-1}\hat{R}^{-1}(t_{n-1})\frac{\partial^{2}}{\partial p_{i}^{2}}\hat{R}(t_{n-1}))\nonumber \\
 &  & \times\cdots(\sum_{i}\int_{0}^{t_{2}}dt_{1}\hat{R}^{-1}(t_{1})\frac{\partial^{2}}{\partial p_{i}^{2}}\hat{R}(t_{1}))]\phi(\vec{x},\vec{p},0)\nonumber \\
 & = & \gamma k_{\mathrm{B}}T(\sum_{i}\hat{R}^{-1}(t)\frac{\partial^{2}}{\partial p_{i}^{2}}\hat{R}(t))\langle\phi(\vec{x},\vec{p},t)\rangle_{\vec{\xi}}.
\end{eqnarray}
Therefore, the observable probability $P(\vec{x},\vec{p},t)=\langle\rho(\vec{x},\vec{p},t)\rangle_{\vec{\xi}}=\hat{R}(t)\langle\phi(\vec{x},\vec{p},t)\rangle_{\vec{\xi}}$
follows the evolution equation 
\begin{eqnarray}
\frac{\partial P}{\partial t} & = & -\hat{L}_{\mathrm{d}}\hat{R}(t)\langle\phi(\vec{x},\vec{p},t)\rangle_{\vec{\xi}}+\hat{R}(t)\frac{\partial}{\partial t}\langle\phi(\vec{x},\vec{p},t)\rangle_{\vec{\xi}}\nonumber \\
 & = & (-\hat{L}_{\mathrm{d}}+\gamma k_{\mathrm{B}}T\sum_{i}\frac{\partial^{2}}{\partial p_{i}^{2}})\hat{R}(t)\langle\phi(\vec{x},\vec{p},t)\rangle_{\vec{\xi}}\nonumber \\
 & = & \sum_{i}[-\frac{\partial}{\partial x_{i}}(\frac{\partial H}{\partial p_{i}}P)+\frac{\partial}{\partial p_{i}}(\frac{\partial H}{\partial x_{i}}P+\gamma\frac{\partial H}{\partial p_{i}}P)+\gamma k_{\mathrm{B}}T\frac{\partial^{2}P}{\partial p_{i}^{2}}],\label{seq:ufflk}
\end{eqnarray}
which is equivalent to Eq.~(4) in the main text.

\section{Shortcuts to isothermality with multiple control parameters\label{SSec-two}}

The framework of the shortcut scheme was originally developed for
systems with single control parameter \citep{Li2017,Albay2019,Li2021}.
To establish a general fromalism, we extend this framework for systems
with multiple control parameters.

In the shortcut scheme, the system evolves according to Eq.~(\ref{seq:ufflk})
with $H=H_{\mathrm{o}}+H_{\mathrm{a}}$. Substituting the instantaneous
equilibrium distribution 
\begin{equation}
P_{\mathrm{eq}}(\vec{x},\vec{p},\vec{\lambda})=\mathrm{e}^{\beta[F(\vec{\lambda})-H_{\mathrm{o}}(\vec{x},\vec{p},\vec{\lambda})]}\label{seq:ipeqd}
\end{equation}
into Eq.~(\ref{seq:ufflk}), we obtain the requirement for the auxillary
Hamiltonian $H_{\mathrm{a}}$ as 
\begin{equation}
\sum_{\mu}(\frac{dF}{d\lambda_{\mu}}-\frac{\partial U_{\mathrm{o}}}{\partial\lambda_{\mu}})\dot{\lambda}_{\text{\ensuremath{\mu}}}=\sum_{i}(\frac{\gamma}{\beta}\frac{\partial^{2}H_{\mathrm{a}}}{\partial p_{i}^{2}}-\gamma p_{i}\frac{\partial H_{\mathrm{a}}}{\partial p_{i}}+\frac{\partial H_{\mathrm{a}}}{\partial p_{i}}\frac{\partial U_{\mathrm{o}}}{\partial x_{i}}-p_{i}\frac{\partial H_{\mathrm{a}}}{\partial x_{i}}).\label{seq:iicondevo}
\end{equation}
The solution for the auxillary Hamiltonian is $H_{\mathrm{a}}(\vec{x},\vec{p},t)=\sum_{\mu}\dot{\lambda}_{\mu}f_{\mu}(\vec{x},\vec{p},\vec{\lambda})$
with $f_{\mu}(\vec{x},\vec{p},\vec{\lambda})$ satisfying 
\begin{equation}
\frac{dF}{d\lambda_{\mu}}-\frac{\partial U_{\mathrm{o}}}{\partial\lambda_{\mu}}=\sum_{i}(\frac{\gamma}{\beta}\frac{\partial^{2}f_{\mu}}{\partial p_{i}^{2}}-\gamma p_{i}\frac{\partial f_{\mu}}{\partial p_{i}}+\frac{\partial f_{\mu}}{\partial p_{i}}\frac{\partial U_{\mathrm{o}}}{\partial x_{i}}-p_{i}\frac{\partial f_{\mu}}{\partial x_{i}}).\label{seq:unauxevof}
\end{equation}
Once the form of the original potential $U_{\mathrm{o}}$ is given,
we can solve Eq.~(\ref{seq:unauxevof}) for the function $f_{\mu}(\vec{x},\vec{p},\vec{\lambda})$.
With the boundary condition 
\begin{equation}
\dot{\vec{\lambda}}(0)=\dot{\vec{\lambda}}(\tau)=0,\label{seq:boundarycon}
\end{equation}
we can realize the shortcut scheme for systems with multi-parameters.

\section{General formalism: The mean work done in the process driven by shortcuts
to isothermality\label{SSec-three}}

In this section, we will derive the work done in the shortcut scheme,
and show its geometric expression.

\subsection{Geometric approach to the irreversible work}

The work done in an individual stochastic trajectory reads \citep{Li2017}
\begin{align}
w[\vec{x}(t),\vec{p}(t)]\equiv & \int_{0}^{\tau}dt\frac{\partial H_{\mathrm{o}}(\vec{x}(t),\vec{p}(t),\vec{\lambda})}{\partial t}+\int_{0}^{\tau}dt\frac{\partial H_{\mathrm{a}}(\vec{x}(t),\vec{p}(t),t)}{\partial t}\nonumber \\
= & \int_{0}^{\tau}dt\frac{\partial H_{\mathrm{o}}(\vec{x}(t),\vec{p}(t),\vec{\lambda})}{\partial t}+\sum_{i}\int_{0}^{\tau}dt(\frac{dH_{\mathrm{a}}}{dt}-\dot{x}_{i}(t)\frac{\partial H_{\mathrm{a}}}{\partial x_{i}}-\dot{p}_{i}(t)\frac{\partial H_{\mathrm{a}}}{\partial p_{i}})\nonumber \\
= & \int_{0}^{\tau}dt\frac{\partial H_{\mathrm{o}}(\vec{x}(t),\vec{p}(t),\vec{\lambda})}{\partial t}-\sum_{i}\int_{0}^{\tau}dt(\dot{x}_{i}(t)\frac{\partial H_{\text{\ensuremath{\mathrm{a}}}}(\vec{x}(t),\vec{p}(t),t)}{\partial x_{i}}+\dot{p}_{i}(t)\frac{\partial H_{\mathrm{a}}(\vec{x}(t),\vec{p}(t),t)}{\partial p_{i}}).\label{seq:trajwork}
\end{align}
In the above derivations, we have used integration by part and considered
the boundary conditions in Eq.~(\ref{seq:boundarycon}). Taking an
ensemble average over the trajectory work $w$ in Eq.~(\ref{seq:trajwork}),
we obtain the mean work as 
\begin{eqnarray}
W & \equiv & \langle w\rangle_{\vec{\xi}}=\iint D[\vec{x}(t)]D[\vec{p}(t)]\mathscr{T}[\vec{x}(t),\vec{p}(t)]w[\vec{x}(t),\vec{p}(t)]\nonumber \\
 & = & \iint D[\vec{x}(t)]D[\vec{p}(t)]\mathscr{T}[\vec{x}(t),\vec{p}(t)]\int_{0}^{\tau}dt\iint d\vec{x}d\vec{p}\delta(\vec{x}-\vec{x}(t))\delta(\vec{p}-\vec{p}(t))\nonumber \\
 &  & \times[\frac{\partial H_{\mathrm{o}}(\vec{x}(t),\vec{p}(t),\vec{\lambda})}{\partial t}-\sum_{i}(\dot{x}_{i}(t)\frac{\partial H_{\text{\ensuremath{\mathrm{a}}}}(\vec{x}(t),\vec{p}(t),t)}{\partial x_{i}}+\dot{p}_{i}(t)\frac{\partial H_{\mathrm{a}}(\vec{x}(t),\vec{p}(t),t)}{\partial p_{i}})]\nonumber \\
 & = & \int_{0}^{\tau}dt\iint d\vec{x}d\vec{p}[\frac{\partial H_{\mathrm{o}}(\vec{x},\vec{p},\vec{\lambda})}{\partial t}\langle\delta(\vec{x}-\vec{x}(t))\delta(\vec{p}-\vec{p}(t))\rangle_{\vec{\xi}}\nonumber \\
 &  & -\sum_{i}(\frac{\partial H_{\text{\ensuremath{\mathrm{a}}}}(\vec{x},\vec{p},t)}{\partial x_{i}}\langle\dot{x}_{i}(t)\delta(\vec{x}-\vec{x}(t))\delta(\vec{p}-\vec{p}(t))\rangle_{\vec{\xi}}+\frac{\partial H_{\mathrm{a}}(\vec{x},\vec{p},t)}{\partial p_{i}}\langle\dot{p}_{i}(t)\delta(\vec{x}-\vec{x}(t))\delta(\vec{p}-\vec{p}(t))\rangle_{\vec{\xi}})]\nonumber \\
 & = & \int_{0}^{\tau}dt\iint d\vec{x}d\vec{p}[\frac{\partial H_{\mathrm{o}}(\vec{x},\vec{p},\vec{\lambda})}{\partial t}P(\vec{x},\vec{p},t)\nonumber \\
 &  & -\sum_{i}(\frac{\partial H_{\text{\ensuremath{\mathrm{a}}}}(\vec{x},\vec{p},t)}{\partial x_{i}}\langle\dot{x}_{i}(t)\rho(\vec{x},\vec{p},t)\rangle_{\vec{\xi}}+\frac{\partial H_{\mathrm{a}}(\vec{x},\vec{p},t)}{\partial p_{i}}\langle\dot{p}_{i}(t)\rho(\vec{x},\vec{p},t)\rangle_{\vec{\xi}})].\label{seq:meanwork}
\end{eqnarray}
In the shortcut scheme, the system keeps in the instantaneous equilibrium
state, $P=P_{\mathrm{eq}}=\exp[\beta(F-H_{\mathrm{o}})],$ which leads
to \citep{Li2017} $\Delta F=\int_{0}^{\tau}dt\iint d\vec{x}d\vec{p}P_{\mathrm{eq}}\partial_{t}H_{\mathrm{o}}.$
The irreversible work $W_{\mathrm{irr}}\equiv W-\Delta F$ then follows
as 
\begin{eqnarray}
W_{\mathrm{irr}} & = & W-\iint d\vec{x}d\vec{p}\int_{0}^{\tau}dt\frac{\partial H_{\mathrm{o}}(\vec{x},\vec{p},\vec{\lambda})}{\partial t}P_{\mathrm{eq}}(\vec{x},\vec{p},t)\nonumber \\
 & = & -\iint d\vec{x}d\vec{p}\sum_{i}\int_{0}^{\tau}dt(\frac{\partial H_{\text{\ensuremath{\mathrm{a}}}}(\vec{x},\vec{p},t)}{\partial x_{i}}\langle\dot{x}_{i}(t)\rho(\vec{x},\vec{p},t)\rangle_{\vec{\xi}}+\frac{\partial H_{\mathrm{a}}(\vec{x},\vec{p},t)}{\partial p_{i}}\langle\dot{p}_{i}(t)\rho(\vec{x},\vec{p},t)\rangle_{\vec{\xi}}).\label{seq:irrworkfirst}
\end{eqnarray}
With Eq. (\ref{seq:ulaneq}), we can calculate $\langle\dot{x}_{i}(t)\rho(\vec{x},\vec{p},t)\rangle_{\vec{\xi}}$
and $\langle\dot{p}_{i}(t)\rho(\vec{x},\vec{p},t)\rangle_{\vec{\xi}}$
as 
\begin{eqnarray}
\langle\dot{x}_{i}(t)\rho(\vec{x},\vec{p},t)\rangle_{\vec{\xi}} & = & \iint D[\vec{x}(t)]D[\vec{p}(t)]\mathscr{T}[\vec{x}(t),\vec{p}(t)]\dot{x}_{i}(t)\delta(\vec{x}-\vec{x}(t))\delta(\vec{p}-\vec{p}(t))\nonumber \\
 & = & \iint D[\vec{x}(t)]D[\vec{p}(t)]\mathscr{T}[\vec{x}(t),\vec{p}(t)]\delta(\vec{x}-\vec{x}(t))\delta(\vec{p}-\vec{p}(t))\frac{\partial H(\vec{x}(t),\vec{p}(t),t)}{\partial p_{i}}\nonumber \\
 & = & \frac{\partial H(\vec{x},\vec{p},t)}{\partial p_{i}}\iint D[\vec{x}(t)]D[\vec{p}(t)]\mathscr{T}[\vec{x}(t),\vec{p}(t)]\delta(\vec{x}-\vec{x}(t))\delta(\vec{p}-\vec{p}(t))\nonumber \\
 & = & \frac{\partial H(\vec{x},\vec{p},t)}{\partial p_{i}}P(\vec{x},\vec{p},t),\label{seq:xaverage}
\end{eqnarray}
and 
\begin{eqnarray}
\langle\dot{p}_{i}(t)\rho(\vec{x},\vec{p},t)\rangle_{\vec{\xi}} & = & \iint D[\vec{x}(t)]D[\vec{p}(t)]\mathscr{T}[\vec{x}(t),\vec{p}(t)]\dot{p}_{i}(t)\delta(\vec{x}-\vec{x}(t))\delta(\vec{p}-\vec{p}(t))\nonumber \\
 & = & \iint D[\vec{x}(t)]D[\vec{p}(t)]\mathscr{T}[\vec{x}(t),\vec{p}(t)]\delta(\vec{x}-\vec{x}(t))\delta(\vec{p}-\vec{p}(t))(-\frac{\partial H(\vec{x}(t),\vec{p}(t),t)}{\partial x_{i}}-\gamma\dot{x}_{i}(t)+\xi_{i}(t))\nonumber \\
 & = & -(\frac{\partial H(\vec{x},\vec{p},t)}{\partial x_{i}}+\gamma\frac{\partial H(\vec{x},\vec{p},t)}{\partial p_{i}})P(\vec{x},\vec{p},t)+\langle\xi_{i}(t)\rho(\vec{x},\vec{p},t)\rangle_{\vec{\xi}}.\label{seq:paverage}
\end{eqnarray}
Since the stochastic force $\vec{\xi}(t)$ commutes with the deterministic
operator $\hat{L}_{\mathrm{d}}(t)$, we have 
\begin{eqnarray}
\langle\xi_{i}(t)\rho(\vec{x},\vec{p},t)\rangle_{\vec{\xi}} & = & \langle\xi_{i}(t)\hat{R}(t)\phi(\vec{x},\vec{p},0)\rangle_{\vec{\xi}}\nonumber \\
 & = & \hat{R}(t)\langle\xi_{i}(t)\phi(\vec{x},\vec{p},t)\rangle_{\vec{\xi}}\nonumber \\
 & = & \hat{R}(t)\langle\xi_{i}(t)[\sum_{n=0}^{\infty}(-1)^{n}\int_{0}^{t}dt_{n}\int_{0}^{t_{n}}dt_{n-1}\cdots\int_{0}^{t_{2}}dt_{1}\hat{O}(t_{n})\hat{O}(t_{n-1})\cdots\hat{O}(t_{1})]\phi(\vec{x},\vec{p},0)\rangle_{\vec{\xi}}\nonumber \\
 & = & \hat{R}(t)[\sum_{n=0}^{\infty}(-1)^{n}\int_{0}^{t}dt_{n}\langle\xi_{i}(t)\hat{O}(t_{n})\rangle_{\vec{\xi}}\int_{0}^{t_{n}}dt_{n-1}\cdots\int_{0}^{t_{2}}dt_{1}\langle\hat{O}(t_{n-1})\cdots\hat{O}(t_{1})\rangle_{\vec{\xi}}]\phi(\vec{x},\vec{p},0)\nonumber \\
 & = & -\hat{R}(t)(\gamma k_{\mathrm{B}}T)\hat{R}^{-1}(t)\frac{\partial}{\partial p_{i}}\hat{R}(t)[\sum_{n=1}^{\infty}(-1)^{n-1}\int_{0}^{t}dt_{n-1}\cdots\int_{0}^{t_{2}}dt_{1}\langle\hat{O}(t_{n-1})\cdots\hat{O}(t_{1})\rangle_{\vec{\xi}}]\phi(\vec{x},\vec{p},0)\nonumber \\
 & = & -\gamma k_{\mathrm{B}}T\frac{\partial}{\partial p_{i}}\hat{R}(t)\langle\phi(\vec{x},\vec{p},t)\rangle_{\vec{\xi}}\nonumber \\
 & = & -\gamma k_{\mathrm{B}}T\frac{\partial}{\partial p_{i}}P(\vec{x},\vec{p},t).\label{seq:xiavera}
\end{eqnarray}
Combining Eqs. (\ref{seq:xaverage}), (\ref{seq:paverage}), and (\ref{seq:xiavera}),
we obtain 
\begin{eqnarray}
W_{\mathrm{irr}} & = & -\sum_{i}\int_{0}^{\tau}dt\iint d\vec{x}d\vec{p}[\frac{\partial H_{\text{\ensuremath{\mathrm{a}}}}}{\partial x_{i}}\frac{\partial H}{\partial p_{i}}P_{\mathrm{eq}}-\frac{\partial H_{\text{\ensuremath{\mathrm{a}}}}}{\partial p_{i}}(\frac{\partial H}{\partial x_{i}}P_{\mathrm{eq}}+\gamma\frac{\partial H}{\partial p_{i}}P_{\mathrm{eq}}+\gamma k_{\mathrm{B}}T\frac{\partial P_{\mathrm{eq}}}{\partial p_{i}})]\nonumber \\
 & = & -\sum_{i}\int_{0}^{\tau}dt\iint d\vec{x}d\vec{p}[\frac{\partial H_{\text{\ensuremath{\mathrm{a}}}}}{\partial x_{i}}\frac{\partial H_{\mathrm{o}}}{\partial p_{i}}-\frac{\partial H_{\text{\ensuremath{\mathrm{a}}}}}{\partial p_{i}}\frac{\partial H_{\mathrm{o}}}{\partial x_{i}}-\gamma(\frac{\partial H_{\mathrm{a}}}{\partial p_{i}})^{2}]P_{\mathrm{eq}}\nonumber \\
 & = & -\sum_{i}\int_{0}^{\tau}dt\iint d\vec{x}d\vec{p}[-\frac{1}{\beta}\frac{\partial H_{\text{\ensuremath{\mathrm{a}}}}}{\partial x_{i}}\frac{\partial P_{\mathrm{eq}}}{\partial p_{i}}+\frac{1}{\beta}\frac{\partial H_{\text{\ensuremath{\mathrm{a}}}}}{\partial p_{i}}\frac{\partial P_{\mathrm{eq}}}{\partial x_{i}}-\gamma(\frac{\partial H_{\mathrm{a}}}{\partial p_{i}})^{2}P_{\mathrm{eq}}]\nonumber \\
 & = & -\sum_{i}\int_{0}^{\tau}dt\iint d\vec{x}d\vec{p}[-\frac{1}{\beta}\frac{\partial}{\partial p_{i}}(\frac{\partial H_{\text{\ensuremath{\mathrm{a}}}}}{\partial x_{i}}P_{\mathrm{eq}})+\frac{1}{\beta}\frac{\partial}{\partial x_{i}}(\frac{\partial H_{\text{\ensuremath{\mathrm{a}}}}}{\partial p_{i}}P_{\mathrm{eq}})-\gamma(\frac{\partial H_{\mathrm{a}}}{\partial p_{i}})^{2}P_{\mathrm{eq}}]\nonumber \\
 & = & \gamma\sum_{i}\int_{0}^{\tau}dt\iint d\vec{x}d\vec{p}\ensuremath{(\frac{\partial H_{\mathrm{a}}}{\partial p_{i}})^{2}}P_{\mathrm{eq}}\nonumber \\
 & = & \gamma\sum_{\mu\nu i}\int_{0}^{\tau}dt\dot{\lambda}_{\mu}\dot{\lambda}_{\nu}\iint d\vec{x}d\vec{p}\ensuremath{\frac{\partial f_{\mu}}{\partial p_{i}}}\ensuremath{\frac{\partial f_{\nu}}{\partial p_{i}}}P_{\mathrm{eq}}\nonumber \\
 & = & \sum_{\mu\nu}\int_{0}^{\tau}dt\dot{\lambda}_{\mu}\dot{\lambda}_{\nu}g_{\mu\nu},\label{seq:underirrcostw}
\end{eqnarray}
with the metric 
\begin{equation}
g_{\mu\nu}=\gamma\sum_{i}\iint d\vec{x}d\vec{p}\ensuremath{\frac{\partial f_{\mu}}{\partial p_{i}}}\ensuremath{\frac{\partial f_{\nu}}{\partial p_{i}}}P_{\mathrm{eq}}\equiv\gamma\sum_{i}\left\langle \ensuremath{\frac{\partial f_{\mu}}{\partial p_{i}}}\ensuremath{\frac{\partial f_{\nu}}{\partial p_{i}}}\right\rangle _{\mathrm{eq}}.\label{seq:undermetric}
\end{equation}
In the derivations of Eq. (\ref{seq:underirrcostw}), we have used
integration by part and assumed that the boundary term 
\[
\sum_{i}(\frac{\partial H_{\text{\ensuremath{\mathrm{a}}}}}{\partial x_{i}}P_{\mathrm{eq}})|{}_{p_{i}=+\infty}=\sum_{i}(\frac{\partial H_{\text{\ensuremath{\mathrm{a}}}}}{\partial x_{i}}P_{\mathrm{eq}})|{}_{p_{i}=-\infty}=0,
\]
and 
\[
\sum_{i}(\frac{\partial H_{\text{\ensuremath{\mathrm{a}}}}}{\partial p_{i}}P_{\mathrm{eq}})|{}_{x_{i}=+\infty}=\sum_{i}(\frac{\partial H_{\text{\ensuremath{\mathrm{a}}}}}{\partial p_{i}}P_{\mathrm{eq}})|{}_{x_{i}=-\infty}=0.
\]
Equation~(\ref{seq:underirrcostw}) is equivlant to Eq.~(7) in the
main text.

\subsection{Positive semi-definteness of the metric $g_{\mu\nu}$}

The positive semi-definiteness of the metric in Eq.~(\ref{seq:undermetric})
are guaranteed by the structure, $g_{\mu\nu}=\gamma\sum_{i}\langle\partial_{p_{i}}f_{\mu}\partial_{p_{i}}f_{\nu}\rangle_{\mathrm{eq}}$.
For any vector $\vec{v}\equiv(v_{1},v_{2},\cdots,v_{M})$, we have
\begin{align}
\vec{v}^{T}g\vec{v} & =\gamma\sum_{\mu\nu}v_{\mu}v_{\nu}\sum_{i}\left\langle \frac{\partial f_{\mu}}{\partial p_{i}}\frac{\partial f_{\nu}}{\partial p_{i}}\right\rangle _{\mathrm{eq}}\nonumber \\
 & =\gamma\sum_{i}\left\langle (\sum_{\mu}v_{\mu}\frac{\partial f_{\mu}}{\partial p_{i}})(\sum_{\nu}v_{\nu}\frac{\partial f_{\nu}}{\partial p_{i}})\right\rangle _{\mathrm{eq}}\nonumber \\
 & =\gamma\sum_{i}\left\langle (\sum_{\mu}v_{\mu}\frac{\partial f_{\mu}}{\partial p_{i}})^{2}\right\rangle _{\mathrm{eq}}\nonumber \\
 & =\gamma\sum_{i}\int d\vec{x}d\vec{p}P_{\mathrm{eq}}(\sum_{\mu}v_{\mu}\frac{\partial f_{\mu}}{\partial p_{i}})^{2}.\label{seq:postitve}
\end{align}
The integrand in Eq.~(\ref{seq:postitve}) is non-negative, which
ensures the non-negativity of $\vec{v}^{T}g\vec{v}$. Therefore, we
prove that the metric in Eq.~(\ref{seq:undermetric}) is positive
semi-definite.

\subsection{The $1/\tau$ scaling of the irreversible work}

It is natural to choose the function form of the protocol $\vec{\lambda}(t)=\vec{\Lambda}(t/\tau)$.
After a change of variable $s\equiv t/\tau$, the control protocol
$\vec{\Lambda}(s)$ is independent of the protocol duration $\tau$.
The irreversible work in Eq. (\ref{seq:underirrcostw}) follows 
\begin{eqnarray*}
W_{\mathrm{irr}} & = & \gamma\sum_{\mu\nu i}\int_{0}^{\tau}dt\dot{\lambda}_{\mu}(t)\dot{\lambda}_{\nu}(t)\iint d\vec{x}d\vec{p}\ensuremath{\frac{\partial f_{\mu}(\vec{x},\vec{p},\vec{\lambda}(t))}{\partial p_{i}}}\ensuremath{\frac{\partial f_{\nu}(\vec{x},\vec{p},\vec{\lambda}(t))}{\partial p_{i}}}P_{\mathrm{eq}}(\vec{x},\vec{p},\vec{\lambda}(t))\\
 & = & \frac{\gamma}{\tau}\sum_{\mu\nu i}\int_{0}^{1}ds\Lambda_{\mu}^{'}(s)\Lambda_{\nu}^{'}(s)\iint d\vec{x}d\vec{p}\ensuremath{\frac{\partial f_{\mu}(\vec{x},\vec{p},\vec{\Lambda}(s))}{\partial p_{i}}}\ensuremath{\frac{\partial f_{\nu}(\vec{x},\vec{p},\vec{\Lambda}(s))}{\partial p_{i}}}P_{\mathrm{eq}}(\vec{x},\vec{p},\vec{\Lambda}(s)),
\end{eqnarray*}
where the prime in $\Lambda_{\mu}^{'}(s)$ represents the derivative
of $\Lambda_{\mu}(s)$ with respective to $s$. The irriversible work
$W_{\mathrm{irr}}$ is inversely proportional to the protocol duration
$\tau$.

\section{General formalism: Solving the geodesic equation with shooting method\label{SSec-fifth}}

According to Eq. (\ref{seq:underirrcostw}), the task of designing
the optimal protocol in the shortcut scheme is converted to finding
the geodesic path in the parameter space. The geodesic path is obtained
by solving the geodesic equation 
\begin{equation}
\ddot{\lambda}_{\mu}+\frac{1}{2}\sum_{\nu\kappa\iota}(g^{-1})_{\iota\mu}(\frac{\partial g_{\iota\nu}}{\partial\lambda_{\kappa}}+\frac{\partial g_{\iota\kappa}}{\partial\lambda_{\nu}}-\frac{\partial g_{\nu\kappa}}{\partial\lambda_{\iota}})\dot{\lambda}_{\nu}\dot{\lambda}_{\kappa}=0,\label{seq:geodesicequa}
\end{equation}
with boundary conditions $\vec{\lambda}(0)=\vec{\lambda}^{0}$, $\vec{\lambda}(\tau)=\vec{\lambda}^{\tau},$
and $\dot{\vec{\lambda}}(0)=\dot{\vec{\lambda}}(\tau)=0$. Shooting
method is one of the popular tools that treats the above two-point
boundary values problem as an initial value problem \citep{Berger2007}.
Specifically, the shooting method solves the initial valve problem
\begin{equation}
\ddot{\lambda}_{\mu}=y_{\mu}(t,\vec{\lambda},\dot{\vec{\lambda}})\equiv\frac{1}{2}\sum_{\nu\kappa\iota}(g^{-1})_{\iota\mu}(\frac{\partial g_{\nu\kappa}}{\partial\lambda_{\iota}}-\frac{\partial g_{\iota\nu}}{\partial\lambda_{\kappa}}-\frac{\partial g_{\iota\kappa}}{\partial\lambda_{\nu}})\dot{\lambda}_{\nu}\dot{\lambda}_{\kappa},\label{seq:vargeodequ}
\end{equation}
with the initial conditions $\vec{\lambda}(0)=\vec{\lambda}^{0}$
and $\dot{\vec{\lambda}}(0+)=\vec{d}$. We remark here that the first
order derivation $\dot{\vec{\lambda}}(0)$ is not continuous at $t=0$,
noticing $\dot{\vec{\lambda}}(0)=0$. The initial rate $\vec{d}$
is updated until the solution of Eq.~(\ref{seq:vargeodequ}) satisfies
the boundary condition $\vec{\lambda}(\tau)=\vec{\lambda}^{\tau}.$
The shooting method can be realized by using the Eular algorithm to
solve Eq. (\ref{seq:vargeodequ}) and Newton's method \citep{Frenkel2001}
to approach the final condition $\vec{\lambda}(\tau)=\vec{\lambda}^{\tau}.$
To update the initial rate $\vec{d}$, we treat the protocol as a
function of the initial rate, i.e., $\vec{\lambda}(t,\vec{d}$), and
define $z_{\mu\nu}(t,\vec{d})\equiv\partial_{d_{\nu}}\lambda_{\mu}(t,\vec{d})$.
At the final time $\tau$, it follows that in Newton's method, the
solution of the equation $\vec{\lambda}^{\tau}=\vec{\lambda}(\tau,\vec{d})$
is approximated as the solution of the equation 
\begin{equation}
\vec{\lambda}^{\tau}\approx\vec{\lambda}(\tau,\vec{d}^{(k)})+z(\tau,\vec{d}^{(k)})(\vec{d}^{(k+1)}-\vec{d}^{(k)}),\label{seq:talerexpansion}
\end{equation}
where $\vec{d}^{(k)}$ represents the current iteration and $\vec{d}^{(k+1)}$
represents the next iteration. Rearranging Eq. (\ref{seq:talerexpansion})
yields 
\begin{equation}
\vec{d}^{(k+1)}=\vec{d}^{(k)}+z^{-1}(\tau,\vec{d}^{(k)})(\vec{\lambda}^{\tau}-\vec{\lambda}(\tau,\vec{d}^{(k)})),\label{seq:iterationeq}
\end{equation}
which gives the process to obtain each new iteration $\vec{d}^{(k+1)}$
from the previous iteration $\vec{d}^{(k)}$. Here, $z^{-1}(\tau,\vec{d}^{(k)})$
in Eq. (\ref{seq:iterationeq}) is obtained by solving the evolution
equation as 
\begin{eqnarray}
\ddot{z}_{\mu\nu}(t,\vec{d}) & = & \frac{\partial\ddot{\lambda}_{\mu}}{\partial d_{\nu}}=\frac{\partial y_{\mu}}{\partial d_{\nu}}\nonumber \\
 & = & \sum_{\kappa}(\frac{\partial y_{\mu}}{\partial\lambda_{\kappa}}\frac{\partial\lambda_{\kappa}}{\partial d_{\nu}}+\frac{\partial y_{\mu}}{\partial\dot{\lambda}_{\kappa}}\frac{\partial\dot{\lambda}_{\kappa}}{\partial d_{\nu}})\nonumber \\
 & = & \sum_{\kappa}(\frac{\partial y_{\mu}}{\partial\lambda_{\kappa}}z_{\kappa\nu}+\frac{\partial y_{\mu}}{\partial\dot{\lambda}_{\kappa}}\dot{z}_{\kappa\nu}),\label{seq:slopder}
\end{eqnarray}
which is derived by taking derivative of Eq. (\ref{seq:vargeodequ})
over $\vec{d}$. The accompanied initial conditions follow as $z_{\mu\nu}(0,\vec{d})=0$
and $\dot{z}_{\text{\ensuremath{\mu\nu}}}(0+,\vec{d})=\text{\ensuremath{\delta_{\text{\ensuremath{\mu\nu}}}}}.$
The shooting method to solve the geodesic equation (\ref{seq:geodesicequa})
is summarized as follows. Firstly, choosing a proper initial rate
$\vec{d}^{(1)}$, we solve Eqs. (\ref{seq:vargeodequ}) and (\ref{seq:slopder})
to obtain the first iteration $\vec{\lambda}(\tau,\vec{d}^{(1)})$
and $z(\tau,\vec{d}^{(1)})$. Secondly, we get the updated rate $\vec{d}^{(2)}$
by using Eq. (\ref{seq:iterationeq}) and repeat the first step to
solve for the next iteration $\vec{\lambda}(\tau,\vec{d}^{(2)})$
and $z(\tau,\vec{d}^{(2)})$. The iterator finally stops in the $k\mathrm{th}$
iteration when $|\vec{\lambda}^{\tau}-\vec{\lambda}(\tau,\vec{d}^{(k)})|<\epsilon$
with $\epsilon$ representing the termination precision. Then, the
solution of the the geodesic equation (\ref{seq:geodesicequa}) is
$\vec{\lambda}(t,\vec{d}^{(k)})$.

\section{Example: Underdamped Brownian motion}

\subsection{The auxiliary Hamiltonian for an one-dimensional system\label{SSec-sixth}}

We consider an underdamped Brownian particle system in an one-dimensional
harmonic potential with the Hamiltonian 
\begin{eqnarray}
H_{\mathrm{o}}(x,p,\vec{\lambda}) & = & \frac{p^{2}}{2}+U_{\mathrm{o}}(x,\vec{\lambda}),\label{seq:auxunderbrow}
\end{eqnarray}
where $U_{\mathrm{o}}(x,\vec{\lambda})=\lambda_{1}(t)x^{2}/2-\lambda_{2}(t)x$
is a controllable potential. The auxiliary Hamiltonian $H_{\mathrm{a}}=H_{\mathrm{a}}(x,p,t)$
follows 
\begin{equation}
(\frac{dF}{d\lambda_{1}}-\frac{\partial U_{\mathrm{o}}}{\partial\lambda_{1}})\dot{\lambda}_{1}+(\frac{dF}{d\lambda_{2}}-\frac{\partial U_{\mathrm{o}}}{\partial\lambda_{2}})\dot{\lambda}_{2}=\frac{\gamma}{\beta}\frac{\partial^{2}H_{\mathrm{a}}}{\partial p^{2}}-\gamma p\frac{\partial H_{\mathrm{a}}}{\partial p}+\frac{\partial U_{\mathrm{o}}}{\partial x}\frac{\partial H_{\mathrm{a}}}{\partial p}-p\frac{\partial H_{\mathrm{a}}}{\partial x}.\label{seq:middleequa}
\end{equation}
The function $f_{1}$ and $f_{2}$ in the auxiliary Hamiltonian $U_{\mathrm{a}}(x,p,t)=\dot{\lambda}_{1}f_{1}(x,p,\lambda_{1},\lambda_{2})+\dot{\lambda}_{2}f_{2}(x,p,\lambda_{1},\lambda_{2})$
satisfy the following equations
\begin{equation}
\frac{\gamma}{\beta}\frac{\partial^{2}f_{1}}{\partial p^{2}}-\gamma p\frac{\partial f_{1}}{\partial p}+\frac{\partial U_{\mathrm{o}}}{\partial x}\frac{\partial f_{1}}{\partial p}-p\frac{\partial f_{1}}{\partial x}=\frac{dF}{d\lambda_{1}}-\frac{\partial U_{\mathrm{o}}}{\partial\lambda_{1}},\label{seq:underauxilf1}
\end{equation}
and 
\begin{equation}
\frac{\gamma}{\beta}\frac{\partial^{2}f_{2}}{\partial p^{2}}-\gamma p\frac{\partial f_{2}}{\partial p}+\frac{\partial U_{\mathrm{o}}}{\partial x}\frac{\partial f_{2}}{\partial p}-p\frac{\partial f_{2}}{\partial x}=\frac{dF}{d\lambda_{2}}-\frac{\partial U_{\mathrm{o}}}{\partial\lambda_{2}}.\label{seq:underauxilf2}
\end{equation}
By assuming that $f_{1}=a_{1}(t)p^{2}+a_{2}(t)xp+a_{3}(t)p+a_{4}(t)x^{2}+a_{5}(t)x$,
we can exactly derive the form 
\begin{equation}
f_{1}(x,p,\lambda_{1},\lambda_{2})=\frac{1}{4\gamma\lambda_{1}}[(p-\gamma x)^{2}+\lambda_{1}x^{2}]-\frac{\lambda_{2}p}{2\lambda_{1}^{2}}+(\frac{\gamma\lambda_{2}}{2\lambda_{1}^{2}}-\frac{\lambda_{2}}{2\gamma\lambda_{1}})x.\label{seq:uauxf1}
\end{equation}
With similar derivations, we can obtain 
\begin{equation}
f_{2}(x,p,\lambda_{1},\lambda_{2})=\frac{p}{\lambda_{1}}-\frac{\gamma x}{\lambda_{1}}.\label{seq:uauxf2}
\end{equation}
Therefore, the auxiliary Hamiltonian takes the form 
\begin{eqnarray}
H_{\mathrm{a}}(x,p,t) & = & \dot{\lambda}_{1}\{\frac{1}{4\gamma\lambda_{1}}[(p-\gamma x)^{2}+\lambda_{1}x^{2}]-\frac{\lambda_{2}p}{2\lambda_{1}^{2}}+(\frac{\gamma\lambda_{2}}{2\lambda_{1}^{2}}-\frac{\lambda_{2}}{2\gamma\lambda_{1}})x\}+\dot{\lambda}_{2}(\frac{p}{\lambda_{1}}-\frac{\gamma x}{\lambda_{1}}).\label{seq:unauxham}
\end{eqnarray}

\subsection{The geodesic protocol for an underdamped Brownian particle system\label{SSec-seventh}}

The metric in Eq.~(\ref{seq:undermetric}) for the current underdamped
Brownian motion takes the form 
\[
g=\left(\begin{array}{cc}
\frac{1}{4\beta\gamma\lambda_{1}^{2}}+\frac{\gamma}{4\beta\lambda_{1}^{3}}+\frac{\gamma\lambda_{2}^{2}}{\lambda_{1}^{4}} & -\frac{\gamma\lambda_{2}}{\lambda_{1}^{3}}\\
-\frac{\gamma\lambda_{2}}{\lambda_{1}^{3}} & \frac{\gamma}{\lambda_{1}^{2}}
\end{array}\right),
\]
which results in the geodesic equations for the minimal work as 
\begin{eqnarray}
\ddot{\lambda}_{1}-\frac{\dot{\lambda}_{1}^{2}(3\gamma^{2}+2\lambda_{1})}{2\lambda_{1}(\gamma^{2}+\lambda_{1})} & = & 0,\nonumber \\
\ddot{\lambda}_{2}-\frac{2\dot{\lambda}_{1}\dot{\lambda}_{2}}{\lambda_{1}}+\frac{\dot{\lambda}_{1}^{2}\lambda_{2}(\gamma^{2}+2\lambda_{1})}{2\lambda_{1}^{2}(\gamma^{2}+\lambda_{1})} & = & 0.\label{seq:twohargeo}
\end{eqnarray}
Two boundary conditions $\vec{\lambda}(0)=\vec{\lambda}^{0}$ and
$\vec{\lambda}(\tau)=\vec{\lambda}^{\tau}$ are accompanied with the
geodesic equation. We first solve Eq.~(\ref{seq:twohargeo}) using
the shooting method mentioned above. The geodesic equation~(\ref{seq:twohargeo})
can be rewritten as 
\begin{eqnarray}
\ddot{\lambda}_{1} & = & y_{1}(t,\vec{\lambda},\dot{\vec{\lambda}})\equiv\frac{\dot{\lambda}_{1}^{2}(2\lambda_{1}+3\gamma^{2})}{2\lambda_{1}(\lambda_{1}+\gamma^{2})},\nonumber \\
\ddot{\lambda}_{2} & = & y_{2}(t,\vec{\lambda},\dot{\vec{\lambda}})\equiv\frac{2\dot{\lambda}_{1}\dot{\lambda}_{2}}{\lambda_{1}}-\frac{\dot{\lambda}_{1}^{2}\lambda_{2}(2\lambda_{1}+\gamma^{2})}{2\lambda_{1}^{2}(\lambda_{1}+\gamma^{2})}.\label{seq:vargeoeq}
\end{eqnarray}
As shown in Sec.~\ref{SSec-fifth}, the algorithm to solve Eq.~(\ref{seq:vargeoeq})
proceeds as follows: 
\begin{algorithm*}
\caption{Shooting method}
\label{algorithm} \KwData{Choose $\dot{\vec{\lambda}}(0+)=\vec{d}(k)$,
with $k=0$; Choose $\delta t$ such that $N\delta t=\tau$ where
$N$ is the number of steps;} \KwResult{Optimal control protocol
$\vec{\lambda}(\tau)$;} \While{$(|\vec{\lambda}^{\tau}-\vec{\lambda}(\tau)|>\mathrm{\epsilon})$}{
$k=k+1$\; $\vec{\lambda}(0)=\vec{\lambda}^{0},$ $\dot{\vec{\lambda}}(0)=\vec{d}(k),$
$z(0)=0,$ $\dot{z}_{11}(0)=\dot{z}_{22}(0)=1,$$\dot{z}_{12}(0)=\dot{z}_{21}(0)=0$\;
\For{$m=0,1,2,\cdots,N-1$}{\For{$\mu=1,2$}{$\lambda_{\mu}((m+1)\delta t)=\lambda_{\mu}(m\delta t)+\dot{\lambda}_{\mu}(m\delta t)\delta t$\;
$\dot{\lambda}_{\mu}((m+1)\delta t)=\dot{\lambda}_{\mu}(m\delta t)+y_{\mu}(m\delta t)\delta t$\;}
\For{$\mu=1,2$}{\For{$\nu=1,2$}{$z_{\mu\nu}((m+1)\delta t)=z_{\mu\nu}(m\delta t)+\dot{z}_{\mu\nu}(m\delta t)\delta t$\;
$\dot{z}_{\mu\nu}((m+1)\delta t)=\dot{z}_{\mu\nu}(m\delta t)+\sum_{\kappa}[\frac{\partial y_{\mu}}{\partial\lambda_{\kappa}}z_{\kappa\nu}+\frac{\partial y_{\mu}}{\partial\dot{\lambda}_{\kappa}}\dot{z}_{\kappa\nu}]\delta t$\;}}}
$d_{1}(k+1)=d_{1}(k)+\frac{(\lambda_{1}^{\tau}-\lambda_{1}(\tau))z_{22}-(\lambda_{2}^{\tau}-\lambda_{2}(\tau))z_{12}}{z_{11}z_{22}-z_{21}z_{12}}$\;
$d_{2}(k+1)=d_{2}(k)+\frac{(\lambda_{2}^{\tau}-\lambda_{2}(\tau))z_{21}-(\lambda_{1}^{\tau}-\lambda_{1}(\tau))z_{11}}{z_{21}z_{12}-z_{11}z_{22}}$\;
}
\end{algorithm*}

\begin{ruledtabular}
In the simulation, we set the parameters as $\vec{\lambda}(0)=(1,1)$,
$\vec{\lambda}(\tau)=(16,2)$, $k_{\mathrm{B}}T=1$, and $\gamma=1$.
The initial rate is chosen as $\vec{d}=(1,1)$ , the operation time
is $\tau=1$ with the time step $\delta t=10^{-3}$, and the termination
precision is set as $\epsilon=10^{-4}.$ The simulation results are
presented in Fig. 2 of the main text. 
\end{ruledtabular}

Fortunately, the geodesic equation~(\ref{seq:twohargeo}) can also
be solved analytically. Substituting the auxiliary Hamiltonian in
Eq.~(\ref{seq:unauxham}) into the irreversible work in Eq.~(\ref{seq:underirrcostw}),
we obtain 
\begin{eqnarray}
\dot{W}_{\mathrm{irr}} & = & \frac{\dot{\lambda}_{1}^{2}}{4\beta\gamma\lambda_{1}^{2}}+\frac{\gamma\dot{\lambda}_{1}^{2}}{4\beta\lambda_{1}^{3}}+\frac{\gamma\dot{\lambda}_{1}^{2}\lambda_{2}^{2}}{\lambda_{1}^{4}}-\frac{2\gamma\dot{\lambda}_{1}\dot{\lambda}_{2}\lambda_{2}}{\lambda_{1}^{3}}+\frac{2\gamma\dot{\lambda}_{2}^{2}}{\lambda_{1}^{2}}\nonumber \\
 & = & \frac{\dot{\lambda}_{1}^{2}(\lambda_{1}+\gamma^{2})}{4\beta\gamma\lambda_{1}^{3}}+\gamma(\frac{\dot{\lambda}_{1}\lambda_{2}}{\lambda_{1}^{2}}-\frac{\dot{\lambda}_{2}}{\lambda_{1}})^{2}\nonumber \\
 & = & \frac{\dot{\lambda}_{1}^{2}(\lambda_{1}+\gamma^{2})}{4\beta\gamma\lambda_{1}^{3}}+\gamma[\frac{d}{dt}(\frac{\lambda_{2}}{\lambda_{1}})]^{2}.\label{seq:irrdissunder}
\end{eqnarray}
We can simplify the expression in Eq. (\ref{seq:irrdissunder}) with
a new set of parameters, 
\begin{equation}
\dot{y}\equiv\dot{\lambda}_{1}\sqrt{\frac{\lambda_{1}+\gamma^{2}}{4\beta\gamma\lambda_{1}^{3}}},\dot{z}\equiv\frac{d}{dt}(\frac{\sqrt{\gamma}\lambda_{2}}{\lambda_{1}}),\label{seq:Eq-munderdamped-xy}
\end{equation}
and the irreversible work follows $\dot{W}_{\mathrm{irr}}=\dot{y}^{2}+\dot{z}^{2},$
indicating a flatten manifold in the geometric space. The corresponding
geodesic equation follows $\ddot{y}=0$ and $\ddot{z}=0$ which gives
\begin{eqnarray}
\dot{\lambda}_{1} & = & \frac{w_{\mathrm{b}}}{\tau}\sqrt{\frac{\lambda_{1}^{3}}{\lambda_{1}+\gamma^{2}}},\nonumber \\
\frac{\lambda_{2}}{\lambda_{1}} & = & m_{\mathrm{b}}t/\tau+n_{\mathrm{b}},\label{seq:Eq-munderdamped-xyeq}
\end{eqnarray}
where the parameters $w_{\mathrm{b}}=-[2\sqrt{1+\gamma^{2}/\lambda_{1}}+\ln(\sqrt{1+\gamma^{2}/\lambda_{1}}-1)-\ln(\sqrt{1+\gamma^{2}/\lambda_{1}}+1)]|_{\lambda_{1}(0)}^{\lambda_{1}(\tau)}$,
$m_{\mathrm{b}}=(\lambda_{2}(\tau)\lambda_{1}(0)-\lambda_{2}(0)\lambda_{1}(\tau))/(\lambda_{1}(\tau)\lambda_{1}(0))$,
and $n_{\mathrm{b}}=\lambda_{2}(0)/\lambda_{1}(0)$. The final geodesic
protocol is shown as Fig. 2 in the main text.

\subsection{The stochastic simulations\label{SSec-eight}}

The motion of the Brownian particle is governed by the Langevin equation
as 
\begin{eqnarray}
\dot{x} & = & \frac{\partial H_{\mathrm{o}}}{\partial p}+\frac{\partial H_{\mathrm{a}}}{\partial p},\nonumber \\
\dot{p} & = & -\frac{\partial H_{\mathrm{o}}}{\partial x}-\frac{\partial H_{\mathrm{a}}}{\partial x}-\gamma\dot{x}+\xi(t),\label{seq:Eq-moverdamped-xlaeq}
\end{eqnarray}
where $\xi$ represents the standard Gaussian white noise satisfying
$\langle\xi(t)\rangle=0$ and $\langle\xi(t)\xi(t')\rangle=2\gamma k_{\mathrm{B}}T\delta(t-t')$.
We introduce the characteristic length $l_{\mathrm{c}}\equiv(k_{\mathrm{B}}T/\lambda_{1}(0))^{1/2}$,
the characteristic times $\tau_{1}=m/\gamma$ and $\tau_{2}=\gamma/\lambda_{1}(0)$
to define the dimensionless coordinate $\tilde{x}\equiv x/l_{\mathrm{c}}$,
momentum $\tilde{p}\equiv p\tau/(ml_{\mathrm{c}})$, time $s\equiv t/\tau$,
and the control protocol $\tilde{\lambda}\equiv\lambda/(kl_{\mathrm{c}}^{2})$.
The dimensionless Langevin equation follows 
\begin{eqnarray}
\tilde{x}' & = & \tilde{p}+\alpha\tilde{\tau}^{2}\frac{\partial\tilde{H}_{\mathrm{a}}}{\partial\tilde{p}},\nonumber \\
\tilde{p}' & = & -\alpha\tilde{\tau}^{2}\frac{\partial\tilde{H}_{\mathrm{o}}}{\partial\tilde{x}}-\alpha\tilde{\tau}^{2}\frac{\partial\tilde{H}_{\mathrm{a}}}{\partial\tilde{x}}-\tilde{\tau}\tilde{x}'+\tilde{\tau}\sqrt{2\alpha\tilde{\tau}}\zeta(s)\text{,}\label{seq:Eq-moverdamped-rxlaeq}
\end{eqnarray}
where $\tilde{\tau}\equiv\tau/\tau_{1}$ and $\alpha\equiv\tau_{1}/\tau_{2}$.
The prime represents the derivative with respective to dimensionless
time $s$. $\zeta(s)$ is a Gaussian white noise satisfying $\langle\zeta(s)\rangle=0$
and $\langle\zeta(s_{1})\zeta(s_{2})\rangle=\delta(s_{1}-s_{2})$.
The Hamiltonian $H_{\mathrm{o}}$ and $H_{\mathrm{a}}$ are rewritten
with the dimensionless parameters as 
\begin{equation}
\tilde{H}_{\mathrm{o}}(\tilde{x},s)=\frac{1}{\alpha\tilde{\tau}^{2}}\frac{\tilde{p}^{2}}{2}+\frac{1}{2}\tilde{\lambda}_{1}\tilde{x}^{2}-\tilde{\lambda}_{2}\tilde{x}\label{seq:Eq-moverdamped-dlessuo}
\end{equation}
and 
\begin{eqnarray}
\tilde{H}_{\mathrm{a}}(\tilde{x},\tilde{p},s) & = & \frac{\tilde{\lambda}'_{1}}{4\tilde{\tau}\tilde{\lambda}_{1}}[\frac{1}{\alpha\tilde{\tau}^{2}}(\tilde{p}-\tilde{\tau}\tilde{x})^{2}+\tilde{\lambda}_{1}\tilde{x}^{2}]\nonumber \\
 &  & -\frac{\tilde{\lambda}'_{1}\tilde{\lambda}_{2}}{2\alpha\tilde{\tau}^{2}\tilde{\lambda}_{1}^{2}}(\tilde{p}+\tilde{\tau}\tilde{x}-\alpha\tilde{\tau}\tilde{\lambda}_{1}\tilde{x})+\tilde{\lambda}'_{2}(\frac{\tilde{p}}{\alpha\tilde{\tau}^{2}\tilde{\lambda}_{1}}-\frac{\tilde{x}}{\alpha\tilde{\tau}\tilde{\lambda}_{1}}).\label{seq:Eq-moverdamped-dlessua}
\end{eqnarray}
We solve the Langevin equation~(\ref{seq:Eq-moverdamped-rxlaeq})
by using the Euler algorithm as 
\begin{eqnarray}
\tilde{x}(s+\delta s) & = & \tilde{x}(s)+\tilde{p}\delta s+\alpha\tilde{\tau}^{2}\frac{\partial\tilde{H}_{\mathrm{a}}}{\partial\tilde{p}}\delta s,\nonumber \\
\tilde{p}(s+\delta s) & = & \tilde{p}(s)-\alpha\tilde{\tau}^{2}\frac{\partial\tilde{H}_{\mathrm{o}}}{\partial\tilde{x}}\delta s-\alpha\tilde{\tau}^{2}\frac{\partial\tilde{H}_{\mathrm{a}}}{\partial\tilde{x}}\delta s-\tilde{\tau}(\tilde{p}+\alpha\tilde{\tau}^{2}\frac{\partial\tilde{H}_{\mathrm{a}}}{\partial\tilde{p}})\delta s+\tilde{\tau}\sqrt{2\alpha\tilde{\tau}\delta s}\theta(s),\label{seq:simulatlange}
\end{eqnarray}
where $\delta s$ is the time step and $\theta(s)$ is a random number
sampled from Gaussian distribution with zero mean and unit variance.
The trajectory work of the system takes 
\begin{eqnarray}
\tilde{w}\equiv\frac{w}{k_{B}T} & = & \int_{0}^{1}(\frac{\partial\tilde{H}_{\mathrm{o}}}{\partial s}+\frac{\partial\tilde{H}_{\mathrm{a}}}{\partial s})ds\nonumber \\
 & \approx & \sum(\frac{\partial\tilde{H}_{\mathrm{o}}}{\partial s}+\frac{\partial\tilde{H}_{\mathrm{a}}}{\partial s})\delta s.\label{seq:Eq-moverdamped-trw}
\end{eqnarray}
In the simulation, we have chosen the parameters $\vec{\lambda}(0)=(1,1)$,
$\vec{\lambda}(\tau)=(16,2)$, $k_{\mathrm{B}}T=1$, $\gamma=1$,
and $m=1$. The mean work is obtained as the ensemble average over
the trajectory work of $10^{5}$ stochastic trajectories. We perform
the simulation for different duration $\tau\in\{0.1,0.5,1.0,1.5,2.0,2.5,3.0\}$.

 \bibliographystyle{unsrt}
\bibliography{ref}